\documentclass[11pt,a4paper]{article}
\pdfoutput=1
\usepackage{jheppub}

\usepackage{multirow, graphicx,amssymb,url,mathrsfs,amsmath}
\usepackage{wrapfig,boxedminipage,subfigure,epsfig}
\usepackage{amsxtra,amstext,latexsym,dsfont,amsfonts}
\usepackage{color}
\usepackage[dvipsnames]{xcolor}
\usepackage{float}
\usepackage{slashed}
\usepackage{calligra}
\DeclareFontShape{T1}{calligra}{m}{n}{<->s*[2.2]callig15}{}
\DeclareMathAlphabet{\mathcalligra}{T1}{calligra}{m}{n}
\usepackage{tikz} 
\usepackage[compat=1.1.0]{tikz-feynman}
\usepackage{feynmf}

\newcommand{\beq}{\begin{equation}}
\newcommand{\eeq}{\end{equation}}
\newcommand{\bea}{\begin{eqnarray}}
\newcommand{\eea}{\end{eqnarray}}

\title{Quantum-gravitational variance of the horizon area}

\title{Quantum uncertainty in the area of a black hole}

\author[a, b]{Maulik Parikh,} 
\author[a]{Jude Pereira}

\affiliation[a]{Department of Physics, Arizona State University, Tempe, Arizona 85287, USA}
\affiliation[b]{Beyond: Center for Fundamental Concepts in Science,
Arizona State University, Tempe, Arizona 85287, USA}

\emailAdd{maulik.parikh@asu.edu}
\emailAdd{jude.pereira@asu.edu}

\abstract{
Quantum fluctuations of the spacetime metric induce an uncertainty in the horizon area of a black hole. Working in linearized quantum gravity, we derive the variance in the area of a four-dimensional Schwarzschild black hole from the renormalized graviton propagator. We find that the standard deviation of the horizon area scales as the product of the Schwarzschild radius and the Planck length. For macroscopic black holes, the quantum uncertainty is therefore enormous in Planck units.
}

\begin{document}
\maketitle
\section{Introduction}
\noindent
The event horizon of a classical black hole is a mathematically precise null hypersurface in spacetime. Correspondingly, the horizon of a Schwarzschild black hole has a definite area. Quantum mechanics of course alters this picture. If any quantum fields are in the Unruh state, a black hole will decay through the probabilistic emission of Hawking radiation -- a process driven by the ability of the horizon to reduce its entropy and thus its area \cite{Parikh:1999mf,Parikh:2004ih}. In this scenario, the expectation value of the area decreases in time. If, however, all quantum fields are in the Hartle-Hawking state, the black hole will remain at equilibrium and the expectation value of the area will stay constant. But even in the Hartle-Hawking state, quantum-gravitational fluctuations of the metric will induce a {\em variance} in the area. The goal of this paper, then, is to calculate the quantum-gravitational variance of the area of a four-dimensional Schwarzschild black hole. We will do that in the context of linearized quantum gravity, for which the fluctuations can be described by a massless helicity-two field living on a Schwarzschild background, and to which the techniques of quantum field theory in curved spacetime are applicable. 


What might we expect to find? A natural guess for the uncertainty (i.e. the standard deviation) might be $\Delta A \sim l_P^2$. Loosely, the rationale would be that the background geometry is a classical solution to Einstein's equations and thus one might expect it to be defined in quantum gravity by a coherent state -- and coherent states are states of minimum uncertainty. This possibility is also indicated by thermality: for a thermal system with $\Delta E \sim T$, a black hole at Hawking temperature $T_H$ would have $\Delta A \sim l_P^2 \Delta S = l_P^2 \Delta E/T_H = l_P^2$. As we will see, however, these naive estimates are not at all what we will obtain by explicit computation.
Rather, we will find that $\Delta A \sim r_H l_P$, where $r_H$ is the Schwarzschild radius. For any macroscopic black hole, this uncertainty is both vastly larger than Planckian scales and disappointingly small on astrophysical scales.

Let us briefly outline the derivation. 
Our first task is to define what we mean by the horizon when the spacetime geometry is fluctuating. In addition, we will need an operator whose expectation value corresponds to the classical area. 
In Section \ref{width}, we argue that linearized quantum gravity offers natural candidates. Since $g_{\theta \theta} = r^2$, we note that the classical horizon area is proportional to the value of $g_{\theta \theta}$ at the horizon. Thus we will take the variance of the horizon area to be proportional to the variance of the local operator $\hat{g}_{\theta \theta}$ evaluated at the coordinate location $r_H$.

To find the variance of a quantum field requires taking the expectation value of the product of operators at a point in spacetime, which is in general divergent. We regulate this via point-splitting, so that $\langle \hat{g}_{\theta \theta}(x)^2 \rangle$ is replaced by $\lim_{x \to x'} \langle \hat{g}_{\theta \theta}(x) \hat{g}_{\theta \theta}(x')\rangle$. This quantity turns out to be proportional to a component of the Feynman propagator for the graviton. Thus to find the regulated variance, we need the position-space graviton propagator on a Schwarzschild background, evaluated near the horizon. This is an arduous calculation but, fortunately, the bulk of the derivation has already been performed by Gaddam and Groenenboem \cite{Gaddam:2020mwe,Gaddam:2020rxb}. They decomposed the metric perturbations into spherical harmonics, imposed Regge-Wheeler gauge, and found the momentum-space propagator of the $(l, m)$th mode in the near-horizon limit. Thus to complete the derivation of the position-space graviton propagator only requires taking the Fourier transform and summing over all modes.

In order to obtain the variance, we need to bring the two points in the propagator together. We can renormalize the resultant divergence by subtracting the graviton propagator in flat space. Somewhat surprisingly, the graviton propagator in flat space has not, to our knowledge, been calculated before in Regge-Wheeler gauge so we follow Gaddam and Groenenboem's methods to obtain the necessary component of the graviton propagator in Minkowski space. (This is not simply a matter of setting the mass to zero in the expression for the black hole graviton propagator because that propagator assumed a near-horizon limit.) We illustrate the method for finding the renormalized position-space propagator for a scalar field, relegating its lengthy graviton counterpart to appendices.

Finally, we take the coincidence limit to obtain the variance of horizon area. We find that uncertainty scales as $r_H l_P$, which is parametrically much larger than what might perhaps have been naively anticipated. We end with some numerical estimates and a discussion of the physical implications for an object free-falling into the black hole. Our derivation relies on perturbative quantum gravity; fluctuations in the spacetime geometry of black hole horizons have also been studied from a holographic perspective \cite{Gukov:2022oed,Freivogel:2024ulb}.


\section{Defining the quantum area of the horizon}\label{width}
\begin{figure}\centering
\includegraphics[width=0.60\textwidth]{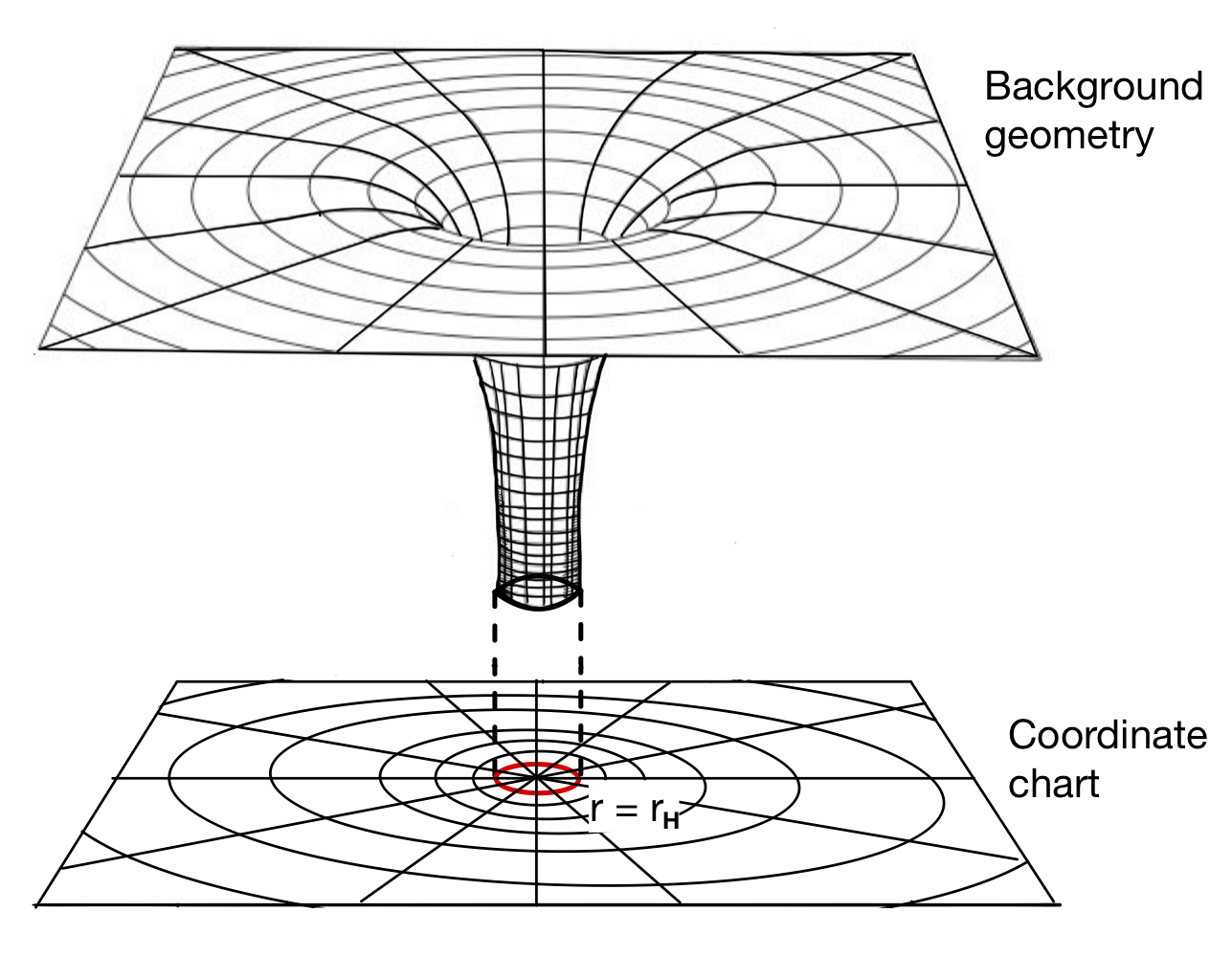}
 \caption{In linearized gravity, the existence of a background geometry can be used to define the variance of the horizon area in a fluctuating geometry. First, we consider a coordinate chart for the background geometry. Keeping the coordinates fixed, we then consider gauge-invariant perturbations of the metric to obtain the variance of the area at the fixed coordinate surface $r = r_H$.}
\label{ProperDistance}
\end{figure}

Consider the line element of a Schwarzschild black hole in the usual coordinates:
\beq
ds^2 = g^{(0)}_{\mu\nu}dx^{\mu}dx^{\nu} = - \Big(1 - \frac{r_H}{r}\Big) dt^2 + \Big(1 - \frac{r_H}{r}\Big)^{-1} dr^2 + r^2 d\theta^2 + r^2 \sin^2\theta d\phi^2 \; .
\eeq
Here $r_H = 2GM$ and the area of the event horizon is
\beq \label{classicalarea}
A = \int_{r = r_H} d \theta d \phi \sqrt{g_{\Omega}} = 4 \pi r_H^2 \; ,
\eeq
where $g_\Omega$ is the determinant of the metric on the two-sphere. The expression \eqref{classicalarea} is only a classical statement; we are interested in calculating its quantum-gravitational fluctuations. More precisely, we wish to calculate the variance of some operator whose expectation value in some appropriate quantum state is given by \eqref{classicalarea}. 
In a full background-independent theory of quantum gravity, this would be a highly non-trivial problem. But in the context of linearized quantum gravity, the presence of the classical background helps us in two ways. First, by quantizing metric perturbations about the background metric, we are able to work in the context of quantum field theory in curved spacetime. And second, the presence of the background allows us to define what we mean by the horizon in a fluctuating geometry, as follows. We pick some coordinates to cover the Schwarzschild geometry, or at least its near-horizon region. By stationarity and spherical symmetry, we can prefentially slice the event horizon into codimension-two sections, and we identify the corresponding surface on the coordinate chart (say $r = r_H$ at some constant $t$). See Figure \ref{ProperDistance}. Keeping the coordinates fixed, we then allow the metric to fluctuate via gauge-invariant perturbations, leading to a variance in the area on that surface. Note that $r_H$ is just a coordinate, labeling the surface of interest on the manifold. When we put the black hole metric on the manifold, the surface has a special significance as the event horizon, but for the Minkowski metric it is just some arbitrary sphere which happens to have the radius $r_H$. 

We also need to decide on an operator that represents the quantum area. A first guess might be to define it using \eqref{classicalarea} in which $g$ would now be a quantum field. However, the presence of the integral would make this operator non-local, and the presence of the determinant would lead to operator-ordering subtleties. In fact, one can verify that the gauge-invariant part of such an operator is identically a constant, even without taking any expectation values (this is easiest to see in Regge-Wheeler gauge). Hence here we will adopt a different approach. We note that, on the horizon, we have
\beq\label{gthetatheta}
g_{\theta\theta}^{(0)}(r_H) = r_H^2 \; .
\eeq
Although $g_{\theta\theta}^{(0)}(t,\vec{x})$ is a local field, by spherical symmetry, its value at any point is related to the area of a two-sphere. Thus in particular, $A = 4 \pi g_{\theta\theta}^{(0)}(r_H)$. In (linearized) quantum gravity, $\hat{g}_{\theta\theta}(x)$ becomes a quantum field. Suppose we have a state for which
\beq\label{expValue}
\langle \hat{g}_{\mu\nu}(x) \rangle = g^{(0)}_{\mu\nu}(x) \; ,
\eeq
We will assume such a state from now on. Then, we can define the expectation value of the area as
\beq\label{gthetatheta}
\text{Exp}[A] \equiv 4 \pi \langle \hat{g}_{\theta\theta}(r_H) \rangle \; .
\eeq
If, in addition, the distribution of fluctuations is also spherically symmetric, then the variance of the area is also proportional to the variance of $\hat{g}_{\theta\theta}$ (analogous to the way that, say in ordinary quantum mechanics, the variance of $\hat{x}$ is proportional to the variance of $\hat{r}$ for a spherically-symmetric distribution). Thus we would expect that -- at least in those quantum states that are invariant under rotations (and thus have a spherically-symmetric distribution of fluctuations) -- the variance of the horizon area would be proportional to the variance of $\hat{g}_{\theta\theta}(x)$, evaluated at the surface $r = r_H$:
\beq\label{gthetatheta}
\text{Var}[A] = 16 \pi^2 \text{Var}[\hat{g}_{\theta\theta}(r_H)] \; .
\eeq
To evaluate this, consider a metric perturbation:
\beq\label{metricfluct}
g_{\mu\nu}(x) = g^{(0)}_{\mu\nu}(x) + \kappa h_{\mu\nu}(x) \; ,
\eeq
where a factor
$\kappa^2 = 8\pi G$ has been included. In linearized gravity, we promote $h$, and therefore $g$, to quantum fields living on a curved manifold with background metric $g^{(0)}$. By Eq.\eqref{expValue}, we
have $\langle \hat{h}_{\mu\nu}(x) \rangle = 0$. 
Then
\beq
\text{Var}[\hat{g}_{\mu\nu}(x)] \equiv \langle \hat{g}_{\mu\nu}(x) \hat{g}_{\mu\nu}(x) \rangle - \big(\langle \hat{g}_{\mu\nu}(x) \rangle\big)^2 
= \kappa^2\langle \hat{h}_{\mu\nu}(x) \hat{h}_{\mu\nu}(x) \rangle \; .
\eeq
A product of two operators taken at the same spacetime point is in general divergent. Thus, to extract a finite quantity, we will need to regularize and renormalize this quantity. Here we will use point-splitting regularization:
\beq
\text{Var}[\hat{g}_{\mu\nu}(x)] = \lim_{\epsilon\rightarrow 0} \kappa^2 \frac{1}{2} \left ( \langle \hat{h}_{\mu\nu}(x) \hat{h}_{\mu\nu}(x + \epsilon) \rangle  + \langle \hat{h}_{\mu\nu}(x + \epsilon) \hat{h}_{\mu\nu}(x) \rangle  \right ) \; .
\eeq
The variance of the metric tensor component is thus equal to the expectation value of the anticommutator of the perturbations. Now, the Feynman propagator obeys the identity
\beq
- i \Delta_F (x,x')
= \frac{1}{2} ( G_R(x,x') + G_A(x,x') ) + \frac{1}{2} G^{(1)}(x,x') \; ,
\eeq
where $G_{R,A}$ are the retarded and advanced propagators and $G^{(1)}$ is the expectation value of the anticommutator. But since the advanced and retarded Green's functions vanish at spacelike separations, our variance can be obtained from the Feynman propagator when $x$ and $x'$ are spacelike-separated. It will be convenient to separate them radially, at the same angle:
\beq
\text{Var}[\hat{g}_{\mu \nu}(x)] = -i \kappa^2 \lim_{\epsilon \to 0} \Delta_{F \, \mu \nu ;\mu \nu} (t,r,\theta,\phi;t,r(1+\epsilon),\theta,\phi) \; .
\eeq
In a Hadamard state, the singular behavior of this quantity is the same as in Minkowski space. Thus, we can obtain a renormalized  quantity by subtracting that expression. Putting everything  together, we have
\beq
\text{Var}[A] = -i 16 \pi^2 \kappa^2 \lim_{\epsilon\to 0} \left (\Delta_{F \, \theta \theta; \theta \theta}^{\text{Sch}} (r_H;r_H + \epsilon')
- \Delta_{F \, \theta \theta; \theta \theta}^{\text{Mink}} (r_H;r_H + \epsilon) \right )
\eeq
Here we have suppressed the other coordinates the propagators depend on. The parameter $\epsilon'$ in the Schwarzschild propagator is chosen such that the proper distance from $r_H$ is $\epsilon$.

\begin{figure}\centering
\includegraphics[width=0.60\textwidth]{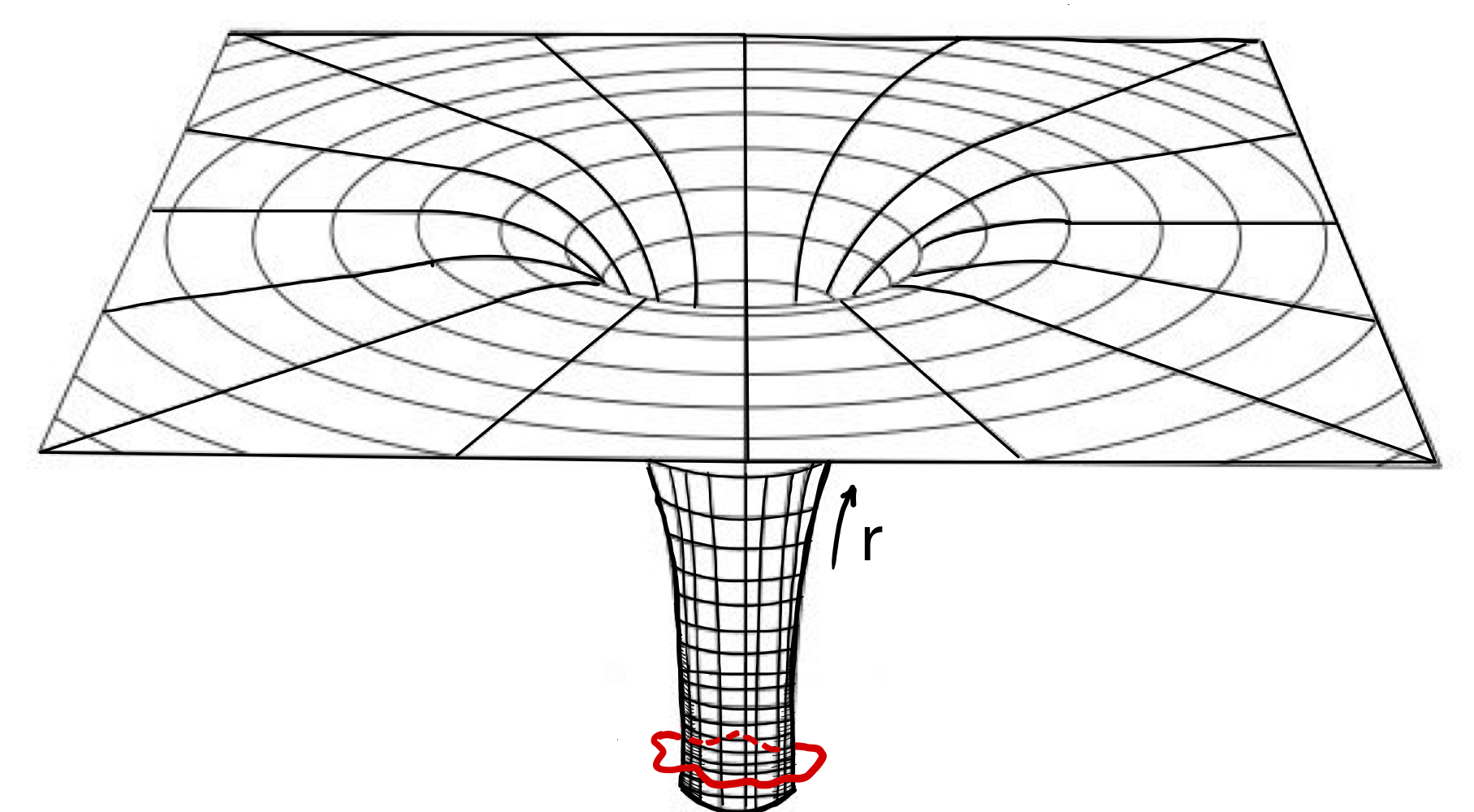}
 \caption{Area changes do not correspond to changes in the radius. The surface of interest is fixed at $r = r_H$ but its area fluctuates because of metric fluctuations over the two-sphere. Geometrically, we are considering fluctuations in the shape of the throat, rather than its depth.}
\label{BHThroat}
\end{figure}

We end this section by clarifying a potential point of confusion. It is incorrect to interpret fluctuations in the area as fluctuations in the horizon radius $r_H$. As we have seen, $r$ is merely a coordinate with $r = r_H$ defining the surface of interest. It does not fluctuate. By contrast, the area of the surface depends on the metric -- and since the metric fluctuates, the area does too. We can view these gauge-invariant  fluctuations of the area as deformations in the shape of a topological two-sphere labeled by a fixed value of $r$. They correspond to deformations in the shape of the throat, as opposed to fluctuations up or down the throat. See Figure \ref{BHThroat}. (There are of course also fluctuations in the radial direction, which we are not considering here.)  In particular, fluctuations in the area should not be confused with a quantum width $\Delta r_H$ of the black hole horizon, or the minimum radius of the stretched horizon \cite{Parikh:1997ma}.


\section{Scalar Propagator}\label{scalar}

We have argued that the variance of the area can be extracted from the $\theta\theta\theta\theta$-component of the renormalized Feynman propagator for the graviton, evaluated on the horizon. 
But before turning to the graviton propagator, let us illustrate the technique by  computing the renormalized Feynman propagator for a scalar. This will allow us to briefly postpone dealing with the various complications that the graviton presents, including the index structure, the coupling between components, and gauge freedom. The idea is to decompose a scalar field living in four dimensions into spherical harmonics whose coefficients constitute an infinite tower of two-dimensional scalar fields. Each of these two-dimensional scalars has a position-space propagator that is readily computed. The desired four-dimensional propagator is then obtained as the sum over all two-dimensional propagators.

Consider then a massless scalar field living in four-dimensional Minkowski space. The action is
\beq
S_{\varphi} = \frac{1}{2}\int d^4x \sqrt{-g} \varphi \Box \varphi
\eeq
The d'Alembertian operator $\Box$ can be expanded as
\beq
\Box = \partial^2 + \frac{2}{r}\partial_r + \frac{1}{r^2}\Delta_{\Omega}
\eeq
where $\partial^2 \equiv -\partial_t^2 + \partial_r^2$.
Expanding the scalar field in spherical harmonics
\beq
\varphi = \sum_{l,m}\varphi_{lm}Y_l^m \; ,
\eeq
using $\Delta_{\Omega}Y_l^m = -l(l+1)Y_l^m$, and integrating over the sphere, we obtain
\beq
S_{\varphi} = - \frac{1}{2}\sum_{l,m}\int d^2x r^2 \varphi_{lm}\Big[-\partial^2 - \frac{2}{r}\partial_r + \frac{l(l+1)}{r^2}\Big]\varphi_{lm}
\eeq
Following \cite{Gaddam:2020mwe}, we make a field redefinition $\Tilde{\varphi} = r\varphi$ to write
\beq
S_{\Tilde{\varphi}} = - \frac{1}{2}\sum_{l,m}\int d^2x \Tilde{\varphi}_{lm}\Big[-\partial^2 + \frac{l(l+1)}{r^2}\Big]\Tilde{\varphi}_{lm}
\eeq
Now we would like to invert the differential operator, giving us the propagator. This can be done by expanding about a fixed timelike hypersurface $r=r_0$. Setting $r = r_0(1+\epsilon)$ and expanding in powers of $\epsilon$, we obtain to leading order in $\epsilon$,
\beq
S_{\Tilde{\varphi}} = - \frac{1}{2}\sum_{l,m}\int d^2x \Tilde{\varphi}_{lm}\Big[-\partial^2 + \frac{l(l+1)}{r_0^2}\Big]\Tilde{\varphi}_{lm}
\eeq
Notice that the leading term is simply a mass term, with effective mass $M^{\varphi}_{\text{Mink}}(l)^2 = l(l + 1)/r_0^2$. Thus, the momentum-space two-dimensional propagator for the $(l,m)$th mode of the scalar field is just
\beq
\mathcal{P}_{\text{Mink}}^{\Tilde{\varphi}}(k) = - \frac{1}{k^2+ l(l + 1)/r_0^2}
\eeq
whose position-space version is obtained via a Fourier transform:
\beq
\mathcal{P}(x;x') = -\int\frac{d^2k}{(2\pi)^2}\frac{1}{k^2+M^{\varphi}_{\text{Mink}}(l)^2}e^{i k_a(x-x')^a}
\eeq
 Here $k^2 = \eta_{ab}k^ak^b = -k_t^2 + k_r^2$ using the 2-dimensional Minkowski metric in $(t,r)$ coordinates. This integral can be readily evaluated by Wick-rotating to Euclidean space.
We find
\beq
\mathcal{P}_{\text{Mink}}^{\Tilde{\varphi}}(s) = \frac{1}{4\pi^2 i}\int_{0}^{\infty}dk \frac{k}{k^2+M^{\varphi}_{\text{Mink}}(l)^2}\int_{0}^{2\pi}d\theta e^{i ks\cos{\theta}}
\eeq
where $s$ is the magnitude of $x-x'$ in Euclidean space.
Using
\beq
e^{i a \cos{\theta}} = \sum_{n=0}^{\infty}i^nJ_n(a)e^{i n \theta}
\eeq
we are left with only $J_0$, after integrating over $\theta$:
\beq
\mathcal{P}_{\text{Mink}}^{\Tilde{\varphi}}(s) = \frac{1}{2\pi i}\int_{0}^{\infty}dk \frac{k}{k^2+M^{\varphi}_{\text{Mink}}(l)^2}J_0(ks)
\eeq
This integral gives
\beq
\mathcal{P}_{\text{Mink}}^{\Tilde{\varphi}}(s) = \frac{1}{2\pi i}K_0(M^{\varphi}_{\text{Mink}}s)
\eeq
where $K_0(x)$ is the modified Bessel function of the second kind. \\ \\
Similarly, the momentum-space two-dimensional scalar propagator for the $(l,m)$th mode in the near-horizon limit is found \cite{Gaddam:2020mwe} to be
\beq
\mathcal{P}_{\text{Sch}}^{\Tilde{\varphi}}(k) = - \frac{1}{k^2+ (l^2 + l + 1)/r_H^2}
\eeq
where $r_H$ is the horizon radius, which corresponds to an effective mass of
$ M^{\varphi}_{\text{Sch}}(l)^2 = (l^2 + l + 1)/r_H^2$. Notice that the effective mass differs from that found in Minkowski space. After a Fourier-transform, we find
\beq
\mathcal{P}_{\text{Sch}}^{\Tilde{\varphi}}(s) = \frac{1}{2\pi i}K_0(M^{\varphi}_{\text{Sch}}s)
\eeq
 One way to renormalize the Schwarzschild propagator is to subtract the corresponding propagator in flat space while setting $r_0 = r_H$. This gives
\beq
\mathcal{P}_{\text{ren}}^{\Tilde{\varphi}}(s) = \frac{1}{2\pi i} \big[K_0(M^{\varphi}_{\text{Sch}}s) - K_0(M^{\varphi}_{\text{Mink}}s)\big]
\eeq
For small $s$, this becomes
\beq\label{approxvalScalar}
\lim_{s\to 0}\mathcal{P}_{\text{ren}}^{\Tilde{\varphi}}(s) = \frac{i}{2\pi}\log\frac{M^{\varphi}_{\text{Sch}}}{M^{\varphi}_{\text{Mink}}}
\eeq
As we will see in Section \ref{result}, the four-dimensional renormalized Feynman propagator is obtained by summing over the $(l, m)$ modes
\beq
\Delta_{F \, \text{ren}}^{\Tilde{\varphi}}(s) = \frac{1}{8\pi^2 i}\sum_{l} (2l + 1) \big[K_0(M^{\varphi}_{\text{Sch}} s) - K_0(M^{\varphi}_{\text{Mink}} s)\big]
\eeq
and then Wick-rotating $s$ back to Lorentzian signature.

\section{Graviton propagator}\label{actionSection}

The graviton propagator in a Schwarzschild background was computed by \cite{Gaddam:2020mwe} in Regge-Wheeler gauge. But since that calculation was performed in the near-horizon limit, we cannot simply take the flat space limit of their result to find the propagator in Minkowski space. We will need the flat-space graviton propagator not in the usual TT gauge but in Regge-Wheeler gauge; surprisingly, we were not able to find this in the existing literature. Here we will calculate the graviton propagator in spherically-symmetric vacuum spacetimes.


\subsection{Einstein-Hilbert action}

To find the graviton propagator, we will need to expand the action to quadratic order in metric perturbations. We take our action to be just the Einstein-Hilbert action: 
\beq
S_{EH}[g_{\mu\nu}] = \frac{1}{2\kappa^2}\int d^4x \sqrt{-g}R \; .
\eeq
In principle, we would also need a ghost action to deal with the gauge freedom. However, it can be shown \cite{Kallosh:2021ors,Kallosh:2021uxa} that this decouples. Next, we use \eqref{metricfluct} to expand $\sqrt{-g}$ and $R_{\mu\nu}$:
\begin{subequations}
    \begin{align}
        \sqrt{-g} &= \sqrt{-g^{(0)}}\Big(1+\frac{1}{2}\kappa h + \mathcal{O}(\kappa^2h^2)\Big) \\
        R_{\mu\nu} &= R_{\mu\nu}^{(0)} + \kappa R_{\mu\nu}^{(1)} + \kappa^2 R^{(2)}_{\mu\nu} + \mathcal{O}(\kappa^3h^3) \; ,
    \end{align} 
\end{subequations}
where we have defined $h = g^{(0)}_{\mu\nu}h^{\mu\nu}$ and the superscripts on $R_{\mu\nu}$ refer to the order in $h$. When $g_{\mu\nu}^{(0)}$ is a vacuum solution to the Einstein equations, both $R_{\mu\nu}^{(0)}$ and the variation of the action to first order in $h_{\mu\nu}$ vanish. Then the action at quadratic order becomes
\beq\label{action1}
S_{EH}^{(2)}[g_{\mu\nu}] = \frac{1}{2}\int d^4x\sqrt{-g^{(0)}}\Big[- \Big(h^{\mu\nu}-\frac{1}{2}g_{(0)}^{\mu\nu}h\Big)R^{(1)}_{\mu\nu} + g_{(0)}^{\mu\nu}R^{(2)}_{\mu\nu}\Big] \; .
\eeq
The last term can be eliminated by discarding boundary terms. 
Then, evaluating the Christoffel symbols to first order, we obtain the quadratic action
\beq\label{action2}
S_{EH}^{(2)}[h_{\mu\nu}] = - \frac{1}{4}\int d^4x\sqrt{-g^{(0)}}\Big(h^{\mu\nu}-\frac{1}{2}g^{\mu\nu}_{(0)}h\Big)(2\nabla^{\sigma}\nabla_{(\mu}h_{\nu)\sigma}-\Box h_{\mu\nu}-\nabla_{\mu}\nabla_{\nu}h) \; ,
\eeq
where tensor indices are raised and lowered using the background metric $g^{(0)}$, which also defines the covariant derivative. To avoid clutter, we have dropped the $(0)$ superscript on the covariant derivatives. Next we need to fix the gauge.

\subsection{Regge-Wheeler Gauge}
Let us write the line element in the form
\beq\label{sphsymmetric}
ds^2 = -2 C(u,v) du dv + r^2(u,v) d\Omega^2 
\eeq
For the Schwarzschild geometry, this turns to the Kruskal-Szekeres coordinates when
\beq\label{SchMetric}
C(u,v) \equiv C(r) = \frac{r_H}{r}\exp{\Big(1- \frac{r}{r_H}\Big)} \qquad uv = 2r_H^2\Big(1-\frac{r}{r_H}\Big)\exp{\Big(\frac{r}{r_H}-1\Big)}
\eeq
Alternatively, when 
\beq
C = 1 \qquad r = \frac{u-v}{\sqrt{2}}
\eeq
the line element describes Minkowski space in lightcone coordinates. Next, we perturb the metric and split the perturbations $h_{\mu\nu}$ into even and odd parity modes:
\beq
h_{\mu\nu} = h^+_{\mu \nu} + h^-_{\mu \nu}
\eeq
where
\beq\begin{aligned}
    h^+_{\mu \nu} &\equiv \sum_{l, m} (h_{lm}^{+})_{\mu\nu} Y_l^m \\ &= \sum_{l, m}\begin{pmatrix}
    H_{uu, lm} & H_{uv, lm} & h_{u,lm}^{+}\partial_{\theta} & h_{u,lm}^{+}\partial_{\phi} \\
    & H_{vv, lm} & h_{v, lm}^{+}\partial_{\theta} & h_{v, lm}^{+}\partial_{\phi} \\
    & & r^2(K_{lm}+G_{lm}\partial_{\theta}^2) & r^2G_{lm} (\partial_{\theta}\partial_{\phi}-\cot{\theta}\partial_{\phi}) \\
    & & & r^2(K_{lm} \sin^2{\theta} + G_{lm}(\partial^2_{\phi}+\sin{\theta}\cos{\theta}\partial_{\theta}))
\end{pmatrix}Y_l^m
\end{aligned}
\eeq
is the even parity perturbation and
\beq\begin{aligned}
    h^-_{\mu \nu} &\equiv \sum_{l, m} (h_{lm}^{-})_{\mu\nu} Y_l^m \\& = \sum_{l, m} \begin{pmatrix}
    0 & 0 & - h^{-}_{u, lm}\csc{\theta}\partial_{\phi} & h_{u,lm}^{-}\sin{\theta}\partial_{\theta} \\
    0 & 0 & - h^{-}_{v, lm}\csc{\theta}\partial_{\phi} & h_{v,lm}^{-}\sin{\theta}\partial_{\theta} \\
    & & h_{\Omega, lm}\csc{\theta}(\partial_{\theta}\partial_{\phi}-\cot{\theta}\partial_{\phi}) & \frac{1}{2}h_{\Omega, lm}\Big(\csc{\theta}\partial_{\phi}^2 + \cos{\theta}\partial_{\theta} - \sin{\theta}\partial^2_{\theta}\Big) \\
    & & & - h_{\Omega, lm} \sin{\theta}(\partial_{\theta}\partial_{\phi}-\cot{\theta}\partial_{\phi})
\end{pmatrix}Y_l^m
\end{aligned}
\eeq
is the odd parity perturbation. Here, for every $l, m$ mode, $H_{uu}, H_{uv}, H_{vv}, h_{u}^{\pm}, h_v^{\pm}, h_{\Omega}, K, G$ (each with implicit $l, m$ subscripts) are ten independent functions of $(u,v)$, comprising the ten independent degrees of the gravitational field.\\ \\
We can use gauge freedom to make four out of the ten independent functions vanish. A convenient gauge, particularly adapted to spherically-symmetric backgrounds, is Regge-Wheeler gauge \cite{Regge:1957td,Zerilli:1970wzz}, for which $G_{lm}, h_{a, lm}^{+}$ and $h_{\Omega, lm}$ are set to zero:
\begin{subequations}
    \begin{align}
        \sum_{l, m} (h_{lm}^{+})_{\mu\nu} Y_l^m  &= \sum_{l, m} \begin{pmatrix}
        H_{uu, lm} & H_{uv, lm} & 0 & 0 \\
        H_{uv, lm} & H_{vv, lm} & 0 & 0 \\
        0 & 0 & r^2K_{lm} & 0 \\
        0 & 0 & 0 & r^2K_{lm}\sin^2{\theta}
        \end{pmatrix}Y_l^m \\
        \sum_{l, m} (h_{lm}^{-})_{\mu\nu} Y_l^m  &= \sum_{l, m} \begin{pmatrix}
        0 & 0 & -h_{u, lm}\csc{\theta}\partial_{\phi} & h_{u, lm}\sin{\theta}\partial_{\theta} \\
        0 & 0 & -h_{v, lm}\csc{\theta}\partial_{\phi} & h_{v, lm}\sin{\theta}\partial_{\theta} \\
         -h_{u, lm}\csc{\theta}\partial_{\phi} & -h_{v, lm}\csc{\theta}\partial_{\phi} & 0 & 0 \\
         h_{u, lm}\sin{\theta}\partial_{\theta} & h_{v, lm}\sin{\theta}\partial_{\theta} & 0 & 0
        \end{pmatrix}Y_l^m
    \end{align}
\end{subequations}
Note that $h^{-}_{\theta\theta}, h^{-}_{\theta\phi}$ and $h^{-}_{\phi\phi}$ vanish in the Regge-Wheeler gauge. Thus the odd parity modes do not contribute to metric perturbations of the sphere. Since our interest here lies in the perturbations of $g_{\theta \theta}$, and since the even and odd parity perturbations decouple in Regge-Wheeler gauge \cite{Gaddam:2020mwe}, we need only consider the even parity perturbations; in what follows, we will drop the $+$ superscript on $h^{+}_{\mu\nu}$. 
Technically, Regge-Wheeler gauge is defined only for the $l \geq 2 $ modes. However, one can define a generalized Regge-Wheeler gauge \cite{Kallosh:2021ors} by making additional gauge choices for the $l = 0, 1$ modes. Making these choices, it can be shown (as reviewed in Appendix \ref{appA}) that the $h_{\theta \theta}$ components for those modes vanish. Thus, for our purposes, the restriction to the $l \geq 2$ modes is sufficient.\\ \\
To summarize, in Regge-Wheeler gauge, the metric perturbations of $g_{\theta \theta}$ can be written as
\beq
h_{\theta \theta} = r^2 K(u, v, \theta, \phi) = \sum_{l = 2}^{\infty}\sum_{m} r^2 K_{lm}(u,v) Y_l^m(\theta, \phi) \; .
\eeq
It can be shown \cite{Martel:2005ir} that, in this gauge, the scalar field $K$ is a gauge-invariant quantity.

\subsection{Deriving the two-dimensional effective action}\label{2daction}
As we saw in the scalar case, the Feynman propagator can be obtained by dimensionally reducing the theory to two dimensions, and summing over the tower of two-dimensional propagators. Following the same approach, we would like to reduce our action \eqref{action2} to two dimensions. After integrating by parts, we find that the Lagrangian for the even parity perturbations can be written as
\beq
L_{\text{even}} = -\frac{1}{8}h^{\mu}_{\ \nu}D^{\nu\ \ \sigma}_{\ \mu\rho}h_{\sigma}^{\ \rho}
\eeq
where we have suppressed the $+$ superscript indicating even parity, and where 
\beq
D^{\nu\ \ \sigma}_{\ \mu\rho} =  \delta^{\sigma}_{\mu}\nabla_{\rho}\nabla^{\nu}+\delta^{\nu}_{\rho}\nabla^{\sigma}\nabla_{\mu}-\delta^{\nu}_{\mu}\nabla^{\sigma}\nabla_{\rho}-\delta^{\sigma}_{\rho}\nabla^{\nu}\nabla_{\mu}+\big(\delta^{\nu}_{\mu}\delta^{\sigma}_{\rho} - \delta^{\nu}_{\rho}\delta^{\sigma}_{\mu}\big) \Box \; .
\eeq
Decomposing the perturbations into spherical harmonics, we can write
\beq\begin{aligned}\label{Gaction}
    D^{\nu\ \ \sigma}_{\ \mu\rho}h_{\sigma}^{\ \rho} =& \sum_{l,m} Y_{l}^{m}D^{\nu\ \ \sigma}_{\ \mu\rho}(h_{lm})_{\sigma}^{\ \rho} + \sum_{l,m} (h_{lm})_{\sigma}^{\ \rho} D^{\nu\ \ \sigma}_{\ \mu\rho} Y_{l}^{m} \\
\end{aligned}\eeq
It is straightforward to check that 
\beq
               D^{\nu\ \ \sigma}_{\ \mu\rho} Y_{l}^{m} = -Y_{l}^{m}\frac{l(l+1)}{2r^2}\Big(\delta^{\nu}_{\mu}\delta^{\sigma}_{\rho} + \delta^{\nu}_{a}\delta^{a}_{\mu}\delta^{\sigma}_{b}\delta^{b}_{\rho} - 2\delta^{\nu}_{\rho}\delta^{\mu}_{\sigma}\Big)
\eeq
where we have used the fact that $\Delta_{\Omega}Y_{l}^{m} = -l(l+1)Y_{l}^{m}$. 
Thus we have 
\beq
S_{\text{even}} = -\frac{1}{8} \sum_{l,m}\sum_{l',m'}\int d\Omega \Bar{Y}_{l'}^{m'}Y_{l}^{m} \int d^2x\ C(u,v)r^2(h_{l'm'})^{\mu}_{\ \nu}\mathcal{D}^{\nu\ \ \sigma}_{\ \mu\rho}(h_{lm})_{\sigma}^{\ \rho}
\eeq
with $C(u,v)$ defined by the metric in \eqref{sphsymmetric}, and
where we have defined
\beq
\mathcal{D}^{\nu\ \ \sigma}_{\ \mu\rho} = D^{\nu\ \ \sigma}_{\ \mu\rho} - \frac{l(l+1)}{2r^2}\Big(\delta^{\nu}_{\mu}\delta^{\sigma}_{\rho} + \delta^{\nu}_{a}\delta^{a}_{\mu}\delta^{\sigma}_{b}\delta^{b}_{\rho} - 2\delta^{\nu}_{\rho}\delta^{\mu}_{\sigma}\Big) \; .
\eeq
Using the orthogonality of the spherical harmonics, we obtain the desired two-dimensional action
\beq
S_{\text{even}} = -\frac{1}{8} \sum_{l,m}\int d^2x\ C(u,v)r^2(h_{lm})^{\mu}_{\ \nu}\mathcal{D}^{\nu\ \ \sigma}_{\ \mu\rho}(h_{lm})_{\sigma}^{\ \rho}
\eeq

\section{Renormalized graviton propagator}\label{MinkProp}

Inserting the even-parity perturbation into the action and reducing to two dimensions, we have 
\beq
S_{\text{even}} = \frac{1}{4}\int d^2x \sqrt{-\text{det}g_{ab}^{(0)}} \Big(\Tilde{H}^{ab}\Delta^{-1}_{abcd}\Tilde{H}^{cd} + \Tilde{H}^{ab}\Delta^{-1}_{L,ab}\Tilde{K} + \Tilde{K}\Delta^{-1}_{R,ab}\Tilde{H}^{ab} + \Tilde{K}\Delta^{-1}\Tilde{K}\Big)
\eeq
where $\text{det}g_{ab}^{(0)}$ is the determinant of the two-dimensional Lorentzian metric. Here, we have absorbed the two-sphere Jacobian factor $r^2$ by redefining $\Tilde{H}_{ab} = r H_{ab}$ and $\Tilde{K} = r K$. The differential operators take the following form:
\begin{subequations}
    \begin{align}
        \Delta^{-1} &= -\Box + F_{a}^{a} \\
        \Delta^{-1}_{R,ab} &= -g_{ab}^{(0)} \Big(\Box - \frac{1}{2} V_c\nabla^c + \frac{1}{4}V_cV^c - F^c_c - \frac{l(l+1)}{2r^2}\Big) + \nabla_a\nabla_b - F_{ab} \\
        \Delta^{-1}_{L,ab} &= -g_{ab}^{(0)} \Big(\Box + \frac{1}{2}V_c\nabla^c - \frac{l(l+1)}{2r^2}\Big) + \nabla_a\nabla_b - F_{ab} \\
        \begin{split}
        \Delta^{-1}_{abcd} &= \frac{1}{2} g^{(0)}_{ac}V_{[b}\nabla_{d]} + \frac{1}{2} g^{(0)}_{bd}V_{[a}\nabla_{c]} + \frac{1}{2} g^{(0)}_{ab} \Big(V_{(c}\nabla_{d)} + 2 F_{cd}\Big) + \frac{1}{2}g^{(0)}_{cd}\Big(-V_{(a}\nabla_{b)} + \frac{1}{2}V_{a}V_{b}\Big) \\& + g^{(0)}_{ab}g^{(0)}_{cd} \Big(\frac{1}{4}R_{2d} + \frac{l(l+1)}{2r^2}\Big) - g^{(0)}_{ac}g^{(0)}_{bd} \Big(\frac{1}{2}R_{2d} + \frac{l(l+1)}{2r^2}\Big) \end{split}
    \end{align}
\end{subequations}
The quantities $V_a$, $F_{ab}$, and $R_{2d}$ depend on the spacetime:
\begin{subequations}
    \begin{align}
        V_a &= 2 \partial_a \log r \\
        F_{ab} &= \frac{1}{r}\nabla_{a}\nabla_{b}r = \frac{1}{2} \nabla_{(a}V_{b)} + \frac{1}{4}V_aV_b \\
        R_{2d} &= -\frac{1}{C} \Box \log{C}
    \end{align}
\end{subequations}
where $C$ is a function of $u$ and $v$.
Now, we are interested in the Feynman propagator for the perturbation field $K$. However, as can be seen from the action, this field is coupled to the $H^{ab}$ field. Thus formally,
we have 
\beq
\mathcal{P}_{K} = \frac{1}{\Delta^{-1}-\Delta^{-1}_{R,ab}\Delta^{abcd}\Delta^{-1}_{L,cd}}
\eeq
Calculating this is a somewhat lengthy calculation, first performed by Gaddam and Groenenboem. They found that, for the $l$th mode, the momentum space graviton propagator in the near horizon limit of the Schwarzschild black hole is
\beq
\mathcal{P}_{\text{Sch}}(k) = - \frac{\lambda+1}{\lambda-3}\frac{1}{k^2 + \mu^2\lambda}
\eeq
where the effective mass is
\beq
M_{\text{Sch}}^2(\lambda) = \mu^2\lambda
\eeq
and $\lambda \equiv l^2 + l + 1$.
In position-space we find
\beq
\mathcal{P}_{\text{Sch}}(s) = \frac{\lambda+1}{\lambda-3}\frac{1}{2\pi i}K_0(M_{\text{Sch}}s)
\eeq
We see from the expression for the black hole graviton propagator that there is a logarithmic UV divergence in the large$-k$ limit. Alternatively, as the two points $x$ and $x'$ are brought close together, the position space propagator diverges because $K_0(M_{\text{Sch}}s)\to\infty$ as $s$ goes to zero. 

Hence to obtain a finite expression for the two-point function, we need to cancel out these divergences in the propagator. Our renormalization procedure consists of subtracting out the corresponding graviton propagator in Minkowski spacetime. To obtain this, we cannot simply take the zero mass limit of the previous result because that was obtained in the near-horizon limit. Thus we have to repeat the calculation of the momentum-space propagator in the vicinity of an arbitrary sphere in Minkowski space. We will of course take the radius of the sphere to be $r_H$. We derive this in the Appendix. We find
\beq
\mathcal{P}(k) = - \frac{\lambda+1}{\lambda-3}\frac{1}{k^2 + \mu^2\Big(\frac{\lambda^2-2\lambda+5}{\lambda-1}\Big)}
\eeq
so that the effective mass is 
\beq
M_{\text{Mink}}^2(\lambda) = \mu^2\frac{\lambda^2-2\lambda+5}{\lambda-1}
\eeq
In position space we have
\beq
\mathcal{P}(s) = \frac{\lambda+1}{\lambda-3}\frac{1}{2\pi i}K_0(M_{\text{Mink}}s)
\eeq
Hence the renormalized propagator in position space is written as
\beq\label{exactval}
\mathcal{P}_{\text{ren}}(s) = \frac{\lambda+1}{\lambda-3}\frac{1}{2\pi i}\big[K_0(M_{\text{Sch}}s)-K_0(M_{\text{Mink}}s)\big]
\eeq
Using the property of modified Bessel functions that $K_0(x) \approx -\log{x}$ for small $x$, it can be checked that the renormalized propagator for each $(l,m)$ mode remains finite in the small $s$ limit
\beq\label{approxvalGrav}
\lim_{s\to 0}\mathcal{P}_{\text{ren}}(s) = \frac{\lambda+1}{\lambda-3}\frac{i}{2\pi}\log\frac{M_{\text{Sch}}(\lambda)}{M_{\text{Mink}}(\lambda)}
\eeq

\section{Determining the variance in the area}\label{result}
We wish to evaluate the variance of $\hat{g}_{\theta \theta}$ at the horizon:
\beq
\text{Var}[g_{\theta\theta}(x_H)] = \lim_{\substack{\epsilon\to 0\\ x \to x_H}} \kappa^2\langle h_{\mu\nu}(x) h_{\mu\nu}(x + \epsilon)\rangle
\eeq
This can be further evaluated as
\begin{subequations}\begin{align}
    \text{Var}[g_{\theta\theta}(x_H)] &= \lim_{x'\to x}\kappa^2\langle h_{\theta\theta}(x) h_{\theta\theta}(x') \rangle_{x,x'\sim x_H} \\
    &= \lim_{x'\to x} \kappa^2r_H^2 \langle\Tilde{K}(x)\Tilde{K}(x')\rangle_{x,x'\sim x_H} \\
    &= \lim_{\substack{\sigma'\to \sigma\\ \Omega'\to\Omega}} \kappa^2r_H^2\sum_{l,m}\sum_{l',m'}\langle\Tilde{K}_{lm}(\sigma)\Tilde{K}_{l'm'}(\sigma')\rangle_{\sigma,\sigma'\sim \sigma_H}Y_l^m(\Omega)\Bar{Y}_{l'}^{m'}(\Omega')\\
    &= \lim_{\substack{s\to 0\\ \Omega'\to\Omega}} -i\kappa^2r_H^2\sum_{l,m}\sum_{l',m'}\mathcal{P}_{\text{Sch}}(s)\delta_{ll'}\delta_{mm'}Y_l^m(\Omega)\Bar{Y}_{l'}^{m'}(\Omega') \\
    &= \lim_{\substack{s\to 0\\ \Omega'\to\Omega}} -i\kappa^2r_H^2\sum_{l,m}\mathcal{P}_{\text{Sch}}(s)Y_l^m(\Omega)\Bar{Y}_{l}^{m}(\Omega')
\end{align}
\end{subequations}
where $\sigma,\sigma'$ have been used to denote the two-dimensional Lorentzian coordinates.
Note that in writing down the last line above, we have imposed the fact that the propagator between distinct $(l,m)$ modes vanishes. This expression can be further simplified using the addition theorem for spherical harmonics
\beq
\sum_{m}Y_l^m(\theta,\phi)\Bar{Y}_{l}^{m}(\theta',\phi') = \frac{2l+1}{4\pi}P_{l}(\cos{\gamma})
\eeq
where $\cos{\gamma}=\cos{\theta}\cos{\theta'}+\cos{(\phi-\phi')}\sin{\theta}\sin{\theta'}$. Substituting this gives
\beq
\text{Var}[g_{\theta\theta}(x_H)] = \lim_{\substack{s\to 0 \\ \gamma\to 0}} \frac{\kappa^2r_H^2}{4\pi i} \sum_{l}(2l+1)P_{l}(\cos{\gamma})\mathcal{P}_{\text{Sch}}(s)
\eeq
As discussed earlier, $\mathcal{P}_{\text{Sch}}(s)$ contains UV divergences which must be suitably renormalized. This can be achieved by replacing $\mathcal{P}_{\text{Sch}}(s)$ with $\mathcal{P}_{\text{ren}}(s)$. Thus we have
\beq
\text{Var}[g_{\theta\theta}(x_H)] = \lim_{\substack{s\to 0 \\ \gamma\to 0}} \frac{\kappa^2r_H^2}{4\pi i} \sum_{l}(2l+1)P_{l}(\cos{\gamma})\mathcal{P}_{\text{ren}}(s)
\eeq
We can simplify this further by choosing $x$ and $x'$ to have the same angular coordinates, thereby setting $\gamma$ to zero in the mode sum:
\beq\label{almostfinal}
\text{Var}[g_{\theta\theta}(x_H)] = \lim_{s\to 0} \frac{\kappa^2r_H^2}{4\pi i} \sum_{l}(2l+1)\mathcal{P}_{\text{ren}}(s)
\eeq
Substituting \eqref{exactval} in \eqref{almostfinal} we have
\beq
\text{Var}[g_{\theta\theta}(x_H)] = \lim_{s\to 0} -\frac{\kappa^2r_H^2}{8\pi^2} \sum_{l=2}^{\infty}(2l+1)\big[K_0(M_{\text{Sch}}s)-K_0(M_{\text{Mink}}s)\big]
\eeq
where $M_{\text{Sch}}$ and $M_{\text{Mink}}$ are the $l$-dependent effective masses of the $l$th spherical harmonic of the metric perturbation $K$.
To evaluate this expression, let us write the sum as
\beq\begin{aligned}
\text{Var}[g_{\theta\theta}(x_H)] = \lim_{s\to 0} -\frac{\kappa^2r_H^2}{8\pi^2} &\sum_{l=2}^{\infty}(2l+1)\big[K_0(M^{\varphi}_{\text{Sch}}s)-K_0(M^{\varphi}_{\text{Mink}}s)\big] \\&+ \frac{\kappa^2r_H^2}{16\pi^2} \sum_{l=2}^{\infty}(2l+1)\frac{4}{\lambda-3}\log\frac{\lambda(\lambda-1)}{\lambda^2-2\lambda+5} 
\end{aligned}\eeq
where $M^{\varphi}_{\text{Sch}}$ and $M^{\varphi}_{\text{Mink}}$ are the effective masses of the scalar field $l$ modes. Here we have used the fact that $M_{\text{Sch}} = M^{\varphi}_{\text{Sch}}$ and approximated $M_{\text{Mink}}$ with $M^{\varphi}_{\text{Mink}}$ in the first sum. With this approximation, which is valid in the large $l$ limit, the first sum is just the the renormalized two-point function for a scalar field, evaluated on the horizon in the coincident limit. That quantity was first computed by Candelas \cite{Candelas:1980zt}, using mode sum techniques. Thus the first term is
\beq
\lim_{s\to 0} \sum_{l=2}^{\infty}(2l+1)\big[K_0(M^{\varphi}_{\text{Sch}}s)-K_0(M^{\varphi}_{\text{Mink}}s)\big] \approx -\frac{1}{6}
\eeq
The original result of \cite{Candelas:1980zt} involved the sum over all the $l$ modes. But it can be checked that the $l=0$ and $l=1$ modes do not contribute, that is, the renormalized propagator corresponding to these modes in \cite{Candelas:1980zt} vanishes. Hence the result of \cite{Candelas:1980zt} can be used for our case, for which the sum over $l$ modes starts at $l=2$.
The second sum can be evaluated by taking a large $l$ limit of the summand as follows
\beq
        \sum_{l=2}^{\infty}(2l+1)\frac{4}{\lambda-3}\log\frac{\lambda(\lambda-1)}{\lambda^2-2\lambda+5}  = \sum_{l=2}^{\infty} \left ( \frac{8}{l^3} - \frac{12}{l^4} + \mathcal{O}(1/l^5) \right) \approx 4(2\zeta(3)+1) - 12 \zeta(4)   
\eeq
Finally, with $\zeta(3) \approx 1.202$ and $\zeta(4) = \pi^4/90$, we have
\beq
             \text{Var}[g_{\theta\theta}(x_H)] \approx \Big(\frac{13}{3} + 8\zeta(3) - \frac{2}{15}\pi^4\Big) \frac{\kappa^2r_H^2}{16\pi^2} \approx \frac{1}{2\pi}r_H^2l_P^2
\eeq
where we have substituted $\kappa^2 = 8\pi l_P^2$.
Thus  the variance of the area of a four-dimensional Schwarzschild black hole in linearized quantum gravity in the Hartle-Hawking state is
\beq
\text{Var}[A] \approx \kappa^2r_H^2 = 8\pi r_H^2l_P^2
\eeq
and thus the standard deviation is
\beq
\Delta A \approx \sqrt{8\pi} r_H l_P
\eeq


\section{Discussion}

We have computed the quantum-gravitational variance of the horizon area for a Schwarzschild black hole, and found that it scales as $r_H l_P$. This scaling has appeared in the literature in various contexts. For example, Marolf \cite{Marolf:2003bb} argued from black hole thermodynamics in $d$ spacetime dimensions that the characteristic scale of the ``quantum width" of the horizon is
\beq
L_c^d \sim \frac{l_P^{d-2}r_H N^{1/2}}{T_H}
\eeq
where $N$ is the number of effective free fields propagating near the black hole and $T_H \sim 1/r_H$ is the Hawking temperature. Setting $d$ to four and the number of effective degrees of freedom to unity, one finds that $L_c \sim r_H^{1/2}l_P^{1/2}$. More recently, Bousso and Penington \cite{Bousso:2023kdj} have argued that the island associated with large-angular momentum Hawking modes of a $d=4$ Schwarzschild black hole extends a distance of order $l_P^{1/2}r_H^{1/2}$ outside the horizon.  Motivated by this, Banks et al \cite{Banks:2024imv} have argued for the breakdown of effective field theory at length scales $l_P^{1/2}r_H^{1/2}$ due to the backreaction from the semiclassical treatment of quantum fields near a black hole.  It is intriguing that the geometric mean of the Schwarzschild radius and the Planck length appears to be a recurring quantity in so many different settings.

Indeed, the other works in the literature are not even calculating the same things we are. To wit, we have computed the variance of the horizon area in the context of linearized quantum gravity. In that theory, the microstates live in the Fock space of the gravitational field. Now, the variance of a quantum operator of course depends on the quantum state. Here we have chosen the Hartle-Hawking vacuum as our quantum state, because this corresponds to a black hole in thermal equilibrium; the Hartle-Hawking state is also the state for which the expectation value of the energy-momentum tensor is regular on both the past and future horizons. The precise relationship between the quantum variance we compute and fluctuations associated with black hole entropy or more fundamental microstate considerations remains unclear, as in the absence of a non-perturbative, holographic theory of quantum gravity, the area does not have a statistical-mechanical interpretation as the entropy. 

Let us also stress that our area fluctuations cannot be interpreted as radial fluctuations. Of course, those are interesting in their own right: heuristically, they could be thought of as the quantum ``width" of the horizon, say, or the distance of the stretched horizon from the true horizon in the context of the membrane paradigm. In principle, the framework employed here could also be utilized to calculate fluctuations in the horizon radius. However, such a computation presents substantial technical challenges, primarily due to the gauge-dependent nature of the radius and the complexity involved in inverting the full set of propagators for tensor and vector perturbations. We leave this more involved calculation to future research.


Let us make some numerical estimates for the characteristic scale $\sqrt{l_P r_H}$. For Sagitarrius A*, the supermassive black hole at the center of the Milky Way, with a mass of $\sim 10^6 M_{\odot}$, the scale corresponds to about an angstrom. That is of course an enormous scale for a quantum-gravity fluctuation, but in a sense it is still likely to be too small to have observational implications. One can also ask whether an infalling observer would be affected by fluctuations on this scale because, at first glance, an observer made up of atoms might perhaps ``feel" fluctuations on the level of an angstrom. However, it must be remembered that the fluctuation in the area is spread over the entire macroscopic area and thus the relative fluctuation in the vicinity of a localized observer is negligible.

Finally, we have considered the quantum-gravitational uncertainty in the horizon of a four-dimensional Schwarzschild black hole. It would be interesting to extend our calculations to other kinds of horizons. 
\\

\noindent
{\bf Acknowledgments}\\
We would like to thank Paul Davies, Ben Freivogel, David Lowe, Erik Verlinde, George Zahariade, and especially Nava Gaddam for helpful discussions. 
JP would like to thank the organizers of the 40th Pacific Coast Gravity meeting at UC Santa Barbara and the 2024 Southwest Strings meeting at Texas A\&M University, where this work was presented. MP is supported in part by Heising-Simons Foundation grant 2021-2818 and by Department of Energy grant DE-SC0019470.
\appendix

\section{Gauge-fixing for $l = 0, 1$ modes}\label{appA}
This can be easily understood from the fact that for the monopole mode $(l=0)$, we have 
\beq
Y_0^0(\theta, \phi) = \frac{1}{2}\frac{1}{\sqrt{\pi}}
\eeq
Thus, the even parity mode becomes
\beq
h^{+}_{\mu\nu, 00} = \begin{pmatrix}
    H_{xx} & H_{xy} & 0 & 0 \\
    & H_{yy} & 0 & 0 \\
    & & r^2K & 0 \\
    & & & r^2K \sin^2{\theta}
\end{pmatrix}Y_0^0
\eeq
while the odd parity mode vanishes identically. Similarly, it can be checked that the diffeomorphisms that define the Regge-Wheeler gauge also vanish. It has been proposed in \cite{Kallosh:2021ors} that a suitable choice of gauge for the monopole mode is to demand $K=0$. The diffeomorphism corresponding to this has been explicitly evaluated in \cite{Gaddam:2022pnb}. The remaining degree of freedom can be fixed by demanding $x_ax^b\epsilon_{bc}h^{ac} = 0$ - this is the form of the generalized Regge-Wheeler gauge defined in \cite{Kallosh:2021ors}. The main takeaway for the present discussion is that for the monopole $l=0$ mode, the $h_{\theta\theta}$ component can be made to vanish through a suitable choice of gauge. Since we are ultimately interested in the metric fluctuations on the sphere, the $l=0$ mode does not contribute to the calculation, and we may neglect it henceforth. 

A similar argument applies to the dipole modes (l=1), where the spherical harmonics are
\beq
Y_{1}^{\pm}(\theta, \phi) = \mp\frac{1}{2}\sqrt{\frac{3}{2\pi}}\sin{\theta}e^{\pm i\phi} \qquad Y_1^0(\theta, \phi) = \frac{1}{2}\sqrt{\frac{3}{\pi}}\cos{\theta}
\eeq
It can be checked that, in this case, the even parity modes become
\beq
h^{+}_{\mu\nu, 1m} = \begin{pmatrix}
    H_{xx, 1m} & H_{xy, 1m} & h_{x,1m}^{+}\partial_{\theta} & h_{x,1m}^{+}\partial_{\phi} \\
    & H_{yy, 1m} & h_{y, 1m}^{+}\partial_{\theta} & h_{y, 1m}^{+}\partial_{\phi} \\
    & & r^2 K_{1m} & 0 \\
    & & & r^2K_{1m} \sin^2{\theta}
\end{pmatrix}Y_1^m
\eeq
and the odd parity modes become
\beq
h^{-}_{\mu\nu, 1m} = \begin{pmatrix}
    0 & 0 & - h^{-}_{x, 1m}\csc{\theta}\partial_{\phi} & h_{x,1m}^{-}\sin{\theta}\partial_{\theta} \\
     & 0 & - h^{-}_{y, 1m}\csc{\theta}\partial_{\phi} & h_{y,1m}^{-}\sin{\theta}\partial_{\theta} \\
    & & 0 & 0 \\
    & & & 0
\end{pmatrix}Y_1^m
\eeq
where we have absorbed $G_{1m}$ into the definition of $K_{1m}$ above. Next, in the even parity modes, we can set $K_{1m} = 0$ and $h_{a, 1m}^{+} = 0$ through the following choice of diffeomorphisms
\begin{subequations}
    \begin{align}
        \xi_a &= \Big(\frac{1}{2}r^2\partial_aK_{1m} - h^{+}_{a,1m}\Big)Y_{1}^m \\
        \xi_A &= -\frac{1}{2}r^2K_{1m}\partial_AY_{1}^m
    \end{align}
\end{subequations}
Thus the even parity mode becomes
\beq
h^{+}_{\mu\nu, 1m} = \begin{pmatrix}
    H_{xx, 1m} & H_{xy, 1m} & 0 & 0 \\
    & H_{yy, 1m} & 0 & 0 \\
    & & 0 & 0 \\
    & & & 0
\end{pmatrix}Y_1^m
\eeq
Once again, the generalized Regge-Wheeler gauge defined in \cite{Kallosh:2021ors} fixes the remaining degree of freedom in the odd parity modes. The details are irrelevant to the present discussion, so we will not present them here. However, the main punchline still holds - the $h_{\theta\theta}$ component can be shown to vanish for the $l=1$ dipole modes through a suitable choice of gauge, and hence, the dipole modes may be neglected in the ensuing discussion. Nevertheless, it is interesting to note that the monopole and the dipole modes, which have been shown not to contain any radiative degrees of freedom \cite{Martel:2005ir} do not contribute to the fluctuations of the horizon.

\section{Deriving the gravity action in Minkowski spacetime}\label{appB}

In this appendix, we derive the final form of the two-dimensional effective action in Minkowski spacetime. The same derivation for Schwarzschild spacetime can be found in \cite{{Gaddam:2020mwe}}.

We start with the Minkowski metric in lightcone-polar coordinates as follows
\beq
ds^2 = -2 dx dy + r^2 d\Omega^2
\eeq
where $x$ and $y$ are lightcone coordinates and the radial coordinate $r$ is defined as 
\beq
r = \frac{x - y}{\sqrt{2}}
\eeq
The derivative of $r$ with respect to the lightcone derivatives in Minkowski spacetime is 
\beq
\partial_ar = \frac{1}{\sqrt{2}}\epsilon_a
\eeq
where we have defined $\epsilon_a = (1 , -1)$.

The non-vanishing Christoffel symbols are 
\begin{subequations}
    \begin{align}
        \Gamma^{\theta}_{\theta x} &= \Gamma^{\theta}_{x\theta} = \Gamma^{\phi}_{\phi x} = \Gamma^{\phi}_{x\phi} = \frac{1}{\sqrt{2}r}\\
        \Gamma^{\theta}_{\theta y} &= \Gamma^{\theta}_{y\theta} = \Gamma^{\phi}_{\phi y} = \Gamma^{\phi}_{y\phi} = -\frac{1}{\sqrt{2}r}\\
        \Gamma^{\phi}_{\theta\phi} &= \Gamma^{\phi}_{\phi\theta} = -\frac{1}{\sin^{2}\theta} \Gamma^{\theta}_{\phi\phi} = \cot{\theta} \\
        \Gamma^{x}_{\theta\theta} &= \frac{1}{\sin^2\theta} \Gamma^x_{\phi\phi} = -\frac{r}{\sqrt{2}}\\
        \Gamma^{y}_{\theta\theta} &= \frac{1}{\sin^2\theta} \Gamma^y_{\phi\phi} = \frac{r}{\sqrt{2}}
    \end{align}
\end{subequations}
It can be checked that all Riemann tensor components vanish as expected since we are in flat spacetime.

Next, we would like to compute all the terms in 
\beq
G^{\nu\ \ \sigma}_{\ \mu\rho}h_{\sigma}^{\ \rho} = \sum_{l,m} Y_{l}^{m}\Big[G^{\nu\ \ \sigma}_{\ \mu\rho} - \frac{l(l+1)}{2r^2}\Big(\delta^{\nu}_{\mu}\delta^{\sigma}_{\rho} + \delta^{\nu}_{a}\delta^{a}_{\mu}\delta^{\sigma}_{b}\delta^{b}_{\rho} - 2\delta^{\nu}_{\rho}\delta^{\mu}_{\sigma}\Big)\Big](h_{lm})_{\sigma}^{\ \rho}
\eeq
The terms in the second parenthesis can be computed easily, so we can focus only on the first term.  Even for the first term, it turns out that since $h^a_{\ b} = H^a_{\ b}$ and $h^A_{\ A} = K$, we only need to compute $G^{A\ \ \sigma}_{\ A\rho}$ and $G^{a\ \ \sigma}_{\ b\rho}$. Hence, we can write
\beq
G^{A\ \ \sigma}_{\ A\rho}h^{\rho}_{\ \sigma} = 2 \nabla^{\rho}\nabla_A h^A_{\ \rho} - 2 \nabla^{\sigma}\nabla_{\rho}h^{\rho}_{\ \sigma} - \nabla_A \nabla^A h + 2 \Box h - \Box h^A_{\ A}
\eeq
In terms of the light-cone fields $H_{ab}$ and $K$, each term in the above expression can be expanded as
\begin{subequations}
    \begin{align}
        \nabla^{\rho}\nabla_Ah^A_{\ \rho} &= \partial_a\big(V^b H^a_{\ b}\big) + \frac{3}{2}V_aV^bH^a_{\ b} - \big(\partial_aV^a\big) K - \frac{3}{2} V^aV_a K \\
        \nabla^{\sigma}\nabla_{\rho}h^{\rho}_{\ \sigma} &= \partial_a\partial^b H^a_{\ b} + V^a \partial_b H_a^{\ b} + \partial_a\big(V^b H^a_{\ b}\big) + V_aV^bH^a_{\ b} - \partial_a\big(V^aK\big) - V_aV^aK \\
        \nabla_A \nabla^A h &= V^a \partial_a H^b_{\ b} + 2V^a \partial_a K \\
        \Box h &= \partial^2H^a_{\ a} + V^b \partial_b H^a_{\ a} + 2 \partial^2K + 2V^b \partial_b K\\
        \Box h^A_{\ A} &= V_a V^b H^a_{\ b} + 2\partial^2K + 2V^a \partial_aK - V_aV^a K
    \end{align}
\end{subequations}
Similarly, we can write 
\beq
G^{a\ \ \sigma}_{\ b\rho\ }h^{\rho}_{\ \sigma} = \nabla^{\rho}\nabla_b h^a_{\ \rho} + \nabla_{\rho}\nabla^a h^{\rho}_{\ b} - \delta^a_b\nabla^{\sigma}\nabla_{\rho}h^{\rho}_{\ \sigma} - \nabla^a\nabla_b h + \delta^a_b \Box h - \Box h^a_{\ b}
\eeq
where the terms can be expanded as 
\begin{subequations}
    \begin{align}
        \nabla^{\rho}\nabla_b h^a_{\ \rho} &= \partial^c\partial_b H^a_{\ c} - \frac{1}{2} V^cV_b H^a_{\ c} + V^c\partial_bH^a_{\ c} - V^a\partial_bK + \frac{1}{2} V^a V_b K \\
        \nabla_{\rho}\nabla^a h^{\rho}_{\ b} &= \partial_c\partial^aH^c_{\ b} - \frac{1}{2} V_c V^a H^c_{\ b} + V_c \partial^a H^c_{\ b} - V^b\partial_a K + \frac{1}{2}V^a V_bK \\
        \nabla^{\sigma}\nabla_{\rho}h^{\rho}_{\ \sigma} &= \partial^a\partial_b H_a^{\ b} + V^a \partial_b H_a^{\ b} + \partial_a\big(V^b H^a_{\ b}\big) + V_aV^b H^a_{\ b} - \partial_a\big(V^a K\big) - V_a V^a K \\
        \nabla^a\nabla_bh &= \partial^a\partial_b H^d_{\ d} + 2 \partial^a\partial_b K \\
        \Box h &= \partial^2H^b_{\ b} + V^a \partial_a H^b_{\ b} + 2 \partial^2 K + 2V^a\partial_aK \\
        \Box h^a_{\ b} &= \Box H^a_{\ b} - \frac{1}{2}V_cV^a H^c_{\ b} - \frac{1}{2} V^c V_b H^a_{\ c} + V^c\partial_c H^a_{\ b} + V^a V_b K
    \end{align}
\end{subequations}
Note that in all the above expressions, we have defined the potential $V_a$ as
\beq
V_a = \Gamma^A_{Aa} = \frac{2}{r} \partial_ar = \frac{\sqrt{2}}{r} \epsilon_a
\eeq
Recalling the form of the action that we had derived at the end of Section \ref{2daction}
\beq
S_{even} = -\frac{1}{8} \sum_{l,m}\int d^2x r^2(h_{lm})^{\mu}_{\ \nu}\mathcal{G}^{\nu\ \ \sigma}_{\ \mu\rho}(h_{lm})_{\sigma}^{\ \rho}
\eeq
This can be further expanded in terms of the light-cone fields $H_{ab}$ and $K$ as
\beq\begin{aligned}
    S_{even} &= -\frac{1}{8} \sum_{l,m}\int d^2x r^2 \Big[H_a^{\ b}\mathcal{G}^{a\ \ d}_{\ bc\ }H^c_{\ d} + H_a^{\ b}\mathcal{G}^{a\ \ A}_{\ bA\ }K + K \mathcal{G}^{A\ \ d}_{\ Ac\ }H^c_{\ d} + K \mathcal{G}^{A\ \ B}_{\ AB\ }K\Big] \\
    &= -\frac{1}{8} \sum_{l,m}\int d^2x r^2 \Big[H^{ab}\mathcal{G}_{abcd}H^{cd} + H^{ab}\mathcal{G}_{L,ab}K + K \mathcal{G}_{R,ab}H^{ab} + K \mathcal{G}K\Big]
\end{aligned}\eeq
Using the identities computed earlier, we can obtain the form of the $\mathcal{G}$ operators as follows
\begin{subequations}
    \begin{align}
        \mathcal{G} &= 2\partial^2 + 2V^a \partial_a \\
        \mathcal{G}_{R, ab} &= 2 \eta_{ab}\Big(\partial^2 + \frac{1}{2}V^c\partial_c - \frac{l(l+1)}{2r^2}\Big) - 2\big(\partial_a\partial_b + V_{(a}\partial_{b)}\big) \\
        \mathcal{G}_{L, ab} &= 2\eta_{ab} \Big(\partial^2 + V^c \partial_c + \frac{1}{2}\partial_c V^c + \frac{1}{2}V_cV^c- \frac{l(l+1)}{2r^2}\Big) - 2\big(\partial_a\partial_b + V_{(a}\partial_{b)}\big) \\
        \begin{split}
            \mathcal{G}_{abcd} &= \eta_{ac} \partial_d\partial_b + \eta_{ac}V_d\partial_b + \eta_{bd} \partial_c\partial_a + \eta_{bd} V_c\partial_a - \eta_{ab} \Big(\partial_c\partial_d + 2V_{(c}\partial_{d)} + V_c V_d + \partial_{(c}V_{d)}\Big) \\& - \eta_{cd}\partial_a\partial_b + \big(\eta_{ab}\eta_{cd} - \eta_{ac}\eta_{bd}\big)\bigg(\partial^2 + V^e\partial_e - \frac{l(l+1)}{r^2}\bigg)
        \end{split}
    \end{align}
\end{subequations}
Further on, it can be checked that in flat Minkowski space 
\beq
\big(\eta_{ac}\partial_d\partial_b + \eta_{bd}\partial_c\partial_a - \eta_{ab}\partial_c\partial_d - \eta_{cd}\partial_a\partial_b + (\eta_{ab}\eta_{cd} - \eta_{ac}\eta_{bd})\partial^2 \big) H^{cd} = 0
\eeq
Using this, the second-order derivative terms in $G_{abcd}$ vanish, and we will be left with 
\beq
    \mathcal{G}_{abcd} = \eta_{ac}V_d\partial_b + \eta_{bd} V_c\partial_a - \eta_{ab} \Big(2V_{(c}\partial_{d)} + V_c V_d + \partial_{(c}V_{d)}\Big) + \big(\eta_{ab}\eta_{cd} - \eta_{ac}\eta_{bd}\big)\bigg(V^e\partial_e - \frac{l(l+1)}{r^2}\bigg)
\eeq
Similarly, the first derivative terms may be rearranged into a symmetric form using
\beq\begin{aligned}
    \Big(\eta_{ac}V_d\partial_b + \eta_{bd} V_c\partial_a &- 2\eta_{ab}V_{(c}\partial_{d)} + \eta_{ab}\eta_{cd}V^e\partial_e - \eta_{ac}\eta_{bd}V^e\partial_e\Big) H^{cd} \\& = \big(\eta_{ac} V_{[d}\partial_{b]} + \eta_{bd}V_{[c}\partial_{a]} - \eta_{ab}V_{(c}\partial_{d)} + \eta_{cd}V_{(a}\partial_{b)}\big)H^{cd}
\end{aligned}\eeq
to write
\beq
    \mathcal{G}_{abcd} = \eta_{ac} V_{[d}\partial_{b]} + \eta_{bd}V_{[c}\partial_{a]} - \eta_{ab}\big(V_{(c}\partial_{d)} + V_cV_d + \partial_{(c}V_{d)}\big) + \eta_{cd}V_{(a}\partial_{b)} - \big(\eta_{ab}\eta_{cd} - \eta_{ac}\eta_{bd}\big)\frac{l(l+1)}{r^2}
\eeq

Next, we would like to redefine the lightcone fields $H_{ab}$ and $K$ as $\Tilde{H}_{ab} = r H_{ab}$ and $\Tilde{K} = r K$. To do so, we need to commute $r$ across the $\mathcal{G}$ operators. We can straightforwardly implement this by defining an alternate derivative
\beq
D_a\phi = \partial_a\phi + \frac{1}{2}V_a\phi = \frac{1}{r}\partial_a(r\phi) 
\eeq
where $\phi$ in the above definition represents the lightcone fields. Using this alternate derivative, we can then redefine the lightcone fields as follows
\beq
r^2 \phi D_a\phi = r \slashed{r} \phi \frac{1}{\slashed{r}}\partial_a(r\phi) = r\phi \partial_a(r\phi) \equiv \Tilde{\phi} \partial_a(\Tilde{\phi})
\eeq
Similarly, it can be checked that
\beq\begin{aligned}
    r^2 \phi D^2\phi &= r^2 \phi D^aD_a \phi = r^2 \phi D^a\Big[\frac{1}{r}\partial_a(r\phi)\Big] \\
    &= r^2 \phi \Big[\partial^a\Big(\frac{1}{r}\partial_a(r\phi)\Big) + \frac{1}{2}V^a\frac{1}{r}\partial_a(r\phi) \Big]\\
    &= r^2 \phi \Big[\frac{1}{r}\partial^2(r\phi) - \frac{1}{r^2}\partial^ar\partial_a(r\phi) + \frac{1}{r^2}\partial^ar\partial_a(r\phi)\Big] \\
    &= r\phi \partial^2(r\phi) = \Tilde{\phi}\partial^2\Tilde{\phi}
\end{aligned}
\eeq
Using $\partial_a\phi = D_a\phi - \frac{1}{2}V_a\phi$, the $\mathcal{G}$ operators can be obtained as follows
\begin{subequations}
    \begin{align}
        \mathcal{G} &= 2 D^2 \\
        \mathcal{G}_{R,ab} &= 2\eta_{ab}\Big(D^2 - \frac{1}{2}V^cD_c + \frac{1}{4}V^cV_c - \frac{l(l+1)}{2r^2}\Big) - 2 D_aD_b \\
        \mathcal{G}_{L,ab} &= 2\eta_{ab} \Big(D^2 + \frac{1}{2}V^cD_c + \frac{1}{4}V_cV^c - \frac{l(l+1)}{2r^2}\Big) - 2D_aD_b \\
        \begin{split}
           \mathcal{G}_{abcd} &= \eta_{ac} V_{[d}D_{b]} + \eta_{bd}V_{[c}D_{a]} - \eta_{ab}\Big(D_{(c}V_{d)} + \frac{1}{2}V_cV_d\Big) + \eta_{cd}\Big(V_{(a}D_{b)}-\frac{1}{2}V_aV_b\Big) \\&- \big(\eta_{ab}\eta_{cd} - \eta_{ac}\eta_{bd}\big)\frac{l(l+1)}{r^2} 
        \end{split}        
    \end{align}
\end{subequations}
Using the above definitions, the final form of the action can be written as 
\beq
S = \sum_{l,m}\frac{1}{4}\int d^2x \Big(\Tilde{H}^{ab}\Delta^{-1}_{abcd}\Tilde{H}^{cd} + \Tilde{H}^{ab}\Delta^{-1}_{L,ab}\Tilde{K} + \Tilde{K}\Delta^{-1}_{R,ab}\Tilde{H}^{ab} + \Tilde{K}\Delta^{-1}\Tilde{K}\Big)
\eeq
where we have defined
\begin{subequations}\begin{align}
           \Delta^{-1} &= -\partial^2 \\
    \Delta^{-1}_{R,ab} &= -\eta_{ab} \Big(\partial^2 -\frac{1}{2}V^c\partial_c +\frac{1}{4}V_cV^c -\frac{l(l+1)}{2r^2}\Big) + \partial_a\partial_b \\
    \Delta^{-1}_{L,ab} &= -\eta_{ab} \Big(\partial^2 +\frac{1}{2}V^c\partial_c -\frac{l(l+1)}{2r^2}\Big) + \partial_a\partial_b \\
    \begin{split}
       \Delta^{-1}_{abcd} &= \frac{1}{2}\eta_{ac}V_{[b}\partial_{d]} + \frac{1}{2}\eta_{bd}V_{[a}\partial_{c]} + \frac{1}{2}\eta_{ab}V_{(c}\partial_{d)} + \frac{1}{2}\eta_{cd}\Big(-V_{(a}\partial_{b)}+\frac{1}{2}V_aV_b\Big) \\&+\frac{l(l+1)}{2r^2}\big(\eta_{ab}\eta_{cd} - \eta_{ac}\eta_{bd}\big)    
    \end{split} 
\end{align}\end{subequations}

\section{Eliminating first derivatives}\label{appC}
We have defined
\begin{subequations}\begin{align}
    \Delta^{-1}_{R,ab} &= -\eta_{ab} \Big[\partial^2 -\mu\partial_r -\frac{1}{2}\mu^2(\lambda-3)\Big] +\partial_a\partial_b \\
    \Delta^{-1}_{L,ab} &= -\eta_{ab} \Big[\partial^2 +\mu\partial_r -\frac{1}{2}\mu^2(\lambda-1)\Big] + \partial_a\partial_b
\end{align}\end{subequations}
The part of the action that comprises these first derivative terms only can be written as
\beq
S_{\partial} = \sum_{l,m}\frac{\mu}{4}\int d^2x \eta_{ab}\big(\Tilde{K}\partial_{r}\Tilde{H}^{ab} -\Tilde{H}^{ab}\partial_{r}\Tilde{K}\big)
\eeq
Since $\Tilde{H}^{ab}$ and $\Tilde{K}$ are functions of light-cone coordinates $(u,v)$, they can be expanded in Fourier modes as follows
\begin{subequations}
    \begin{align}
        \Tilde{H}^{ab}(x) &= \frac{1}{(2\pi)^2}\int d^2k e^{i k_ax^a} \Tilde{H}^{ab}(k) \\
        \Tilde{K}(x) &= \frac{1}{(2\pi)^2}\int d^2k' e^{i k'_ax^a}\Tilde{K}(k')
    \end{align}
\end{subequations}
Substituting back in the action, we obtain
\beq
S_{\partial} = \sum_{l,m}\frac{\mu}{4}\frac{1}{(2\pi)^4}\int d^2x \eta_{ab}\int d^2k d^2k'\big[e^{i (k_a - k'_a)x^a}\Tilde{K}(k')k_{r}\Tilde{H}^{ab}(k) -e^{i (k'_a - k_a)x^a}\Tilde{H}^{ab}(k)k'_{r}\Tilde{K}(k')\big]
\eeq
Performing the position space integral and evaluating the delta function, we get
\beq
S_{\partial} = \sum_{l,m}\frac{\mu}{4}\frac{1}{(2\pi)^2} \eta_{ab}\int d^2k \big[\Tilde{K}(k)k_{r}\Tilde{H}^{ab}(k) - \Tilde{H}^{ab}(k)k_{r}\Tilde{K}(k)\big] = 0
\eeq
Hence the first derivative terms in $\Delta^{-1}_{L, ab}$ and $\Delta^{-1}_{R, ab}$ do not contribute to the action and can be discarded.

\section{Obtaining the $K$ propagator in Minkowski space}\label{appD}

Here we will repeat the steps of Gaddam and Groenenboem's calculation for the $K$ part of the graviton propagator in the case of Minkowski spacetime.
\beq\label{action}
S_{\text{even}} = \sum_{l,m}\frac{1}{4}\int d^2x \Big(\Tilde{H}^{ab}\Delta^{-1}_{abcd}\Tilde{H}^{cd} + \Tilde{H}^{ab}\Delta^{-1}_{L,ab}\Tilde{K} + \Tilde{K}\Delta^{-1}_{R,ab}\Tilde{H}^{ab} + \Tilde{K}\Delta^{-1}\Tilde{K}\Big)
\eeq
The potentials take the form
\beq\begin{aligned}
    V_a &= 2 \partial_a\log{r} = \frac{2}{r}\partial_ar \\
    F_{ab} &= \frac{1}{r} \partial_a\partial_br \implies F_{ab} = 0 \\
    R_{2d} &= 0
\end{aligned}\eeq

Under this choice of potentials, the $\Delta^{-1}$ operators are defined as
\begin{subequations}\begin{align}
           \Delta^{-1} &= -\partial^2 \\
    \Delta^{-1}_{R,ab} &= -\eta_{ab} \Big(\partial^2 -\frac{1}{2}V_c\partial^c +\frac{1}{4}V_cV^c -\frac{l(l+1)}{2r^2}\Big) + \partial_a\partial_b \label{DeltainvR} \\
    \Delta^{-1}_{L,ab} &= -\eta_{ab} \Big(\partial^2 +\frac{1}{2}V_c\partial^c -\frac{l(l+1)}{2r^2}\Big) + \partial_a\partial_b \label{DeltainvL}\\
    \begin{split}\label{invdeltaabcd}
       \Delta^{-1}_{abcd} &= \frac{1}{2}\eta_{ac}V_{[b}\partial_{d]} + \frac{1}{2}\eta_{bd}V_{[a}\partial_{c]} + \frac{1}{2}\eta_{ab}V_{(c}\partial_{d)} + \frac{1}{2}\eta_{cd}\Big(-V_{(a}\partial_{b)}+\frac{1}{2}V_aV_b\Big) \\&+\frac{l(l+1)}{2r^2}\big(\eta_{ab}\eta_{cd} - \eta_{ac}\eta_{bd}\big)    
    \end{split} 
\end{align}\end{subequations}
Further on, substituting $V_cV^c = 4/r^2$ and $V_c\partial^c = (2/r)\partial_r$ in \eqref{DeltainvR} and \eqref{DeltainvL} we obtain
\begin{subequations}\begin{align}
    \Delta^{-1}_{R,ab} &= -\eta_{ab} \Big(\partial^2 -\frac{1}{r}\partial_r -\frac{1}{2r^2}(\lambda-3)\Big) +\partial_a\partial_b \\
    \Delta^{-1}_{L,ab} &= -\eta_{ab} \Big(\partial^2 +\frac{1}{r}\partial_r -\frac{1}{2r^2}(\lambda-1) \Big) + \partial_a\partial_b
\end{align}\end{subequations}
where we have defined $\lambda = l^2 + l + 1$. Further on, the potential terms can be converted to mass terms by setting $r = r_H(1 + \epsilon)$. Since $\epsilon \ll 1$, we can expand in powers of $\epsilon$ and hence to leading order in $\epsilon$, we obtain
\begin{subequations}\begin{align}
    \Delta^{-1}_{R,ab} &= -\eta_{ab} \Big(\partial^2 - \mu \partial_r -\frac{1}{2}\mu^2(\lambda-3)\Big) +\partial_a\partial_b \\
    \Delta^{-1}_{L,ab} &= -\eta_{ab} \Big(\partial^2 + \mu \partial_r -\frac{1}{2}\mu^2(\lambda-1) \Big) + \partial_a\partial_b
\end{align}\end{subequations}
where we have defined $\mu = 1/r_H$. In Appendix \ref{appC}, we have shown that since the first derivative terms in $\Delta^{-1}_{R,ab}$ and  $\Delta^{-1}_{L,ab}$ have opposite signs, they cancel out in the action and hence can be discarded. 

The equations of motion can be derived using the action \eqref{action}
\begin{subequations}\begin{align}
    \Delta^{-1}_{abcd}\Tilde{H}^{cd} + \Delta^{-1}_{L,ab}\Tilde{K} &= 0\\
    \Delta^{-1}_{R,ab}\Tilde{H}^{ab} + \Delta^{-1}\Tilde{K} &= 0
\end{align}\end{subequations}
Writing in matrix form, we have
\beq
\begin{pmatrix}
    \Delta^{-1}_{abcd} & \Delta^{-1}_{L,ab} \\
    \Delta^{-1}_{R,cd} & \Delta^{-1}
\end{pmatrix}
\begin{pmatrix}
    \Tilde{H}^{cd} \\
    \Tilde{K}
\end{pmatrix}
= 0
\eeq
We define the Green's function propagator as
\beq
\begin{pmatrix}
    \Delta^{-1}_{abcd} & \Delta^{-1}_{L,ab} \\
    \Delta^{-1}_{R,cd} & \Delta
\end{pmatrix}
\begin{pmatrix}
    \mathcal{P}^{cdef} & \mathcal{P}^{cd}_R \\
    \mathcal{P}^{ef}_L & \mathcal{P}
\end{pmatrix}
= 
\begin{pmatrix}
    \delta^{ef}_{ab} & 0 \\
    0 & 1
\end{pmatrix}
\delta^{(2)}(x^a - x'^a)
\eeq
This can be explicitly written in terms of the following equations:
\begin{subequations}\begin{align}
    \Delta^{-1}_{abcd}\mathcal{P}^{cdef} + \Delta^{-1}_{L,ab}\mathcal{P}^{ef}_L &= \delta^{ef}_{ab}\delta^{(2)}(x^a - x'^a)\\
    \Delta^{-1}_{abcd}\mathcal{P}^{cd}_R + \Delta^{-1}_{L,ab}\mathcal{P} &= 0 \\
    \Delta^{-1}_{R,ab}\mathcal{P}^{abcd} + \Delta^{-1}\mathcal{P}^{cd}_L &= 0 \\
    \Delta^{-1}_{R,ab}\mathcal{P}^{ab}_R + \Delta^{-1}\mathcal{P} &= \delta^{(2)}(x^a - x'^a)
\end{align}\end{subequations}
The inverse of the operators is defined as follows
\begin{subequations}
    \begin{align}
        \label{deltaabcd}
        \Delta^{-1}_{abcd}\Delta^{cdef} &= \delta^{ef}_{ab}\delta^{(2)}(x^a - x'^a) \\
        \Delta^{-1}_{ab}\Delta^{bc} &= \delta^c_a\delta^{(2)}(x^a - x'^a) \\
        \Delta^{-1}\Delta &= \delta^{(2)}(x^a - x'^a)
    \end{align}
\end{subequations}
To find the inverse of $\Delta^{-1}$, we substitute the following Fourier expansions
\beq\begin{aligned}
    \Delta(x;x') &= \frac{1}{(2\pi)^2}\int d^2k e^{i k_a(x-x')^a} \Delta(k) \\
\delta^{(2)}(x^a - x'^a) &= \frac{1}{(2\pi)^2}\int d^2k e^{i k_a(x-x')^a}
\end{aligned}
\eeq
This gives $\Delta(k) = \frac{1}{k^2}$ where $k^2 = \eta^{ab}k_ak_b$.\\ 
To find the inverse of $\Delta^{-1}_{abcd}$, we need to solve \eqref{deltaabcd}. We work with the form of $\Delta^{-1}_{abcd}$ as in \eqref{invdeltaabcd}. Since the RHS is a delta function distribution, we can interpret the LHS of \eqref{deltaabcd} as being under an integral. In that case, using integration by parts, one can replace the following terms
\begin{subequations}
    \begin{align}
        \eta_{ac}V_{[b}\partial_{d]}\Delta^{cdef}(x;x') &\rightarrow -\eta_{ac}(\partial_{[d}V_{b]})\Delta^{cdef} \\
        \eta_{bd}V_{[a}\partial_{c]}\Delta^{cdef}(x;x') &\rightarrow -\eta_{bd}(\partial_{[c}V_{a]})\Delta^{cdef} \\
        \eta_{ab}V_{(c}\partial_{d)}\Delta^{cdef}(x;x') &\rightarrow -\eta_{ab}(\partial_{(d}V_{c)})\Delta^{cdef} \\
        \eta_{cd}V_{(a}\partial_{b)}\Delta^{cdef}(x;x') &\rightarrow -\eta_{cd}(\partial_{(b}V_{a)})\Delta^{cdef}
    \end{align}
\end{subequations}
Further on, we have
\beq
V_b = \frac{2}{r}\partial_br = \frac{\sqrt{2}}{r}\epsilon_b
\eeq
where we have defined $\epsilon_b = (1, -1)^{T}$. Using this, we can evaluate
\beq\begin{aligned}
    \partial_aV_b &= -\frac{1}{r^2}\epsilon_a\epsilon_b \\
    \partial_{[a}V_{b]} &= 0\\
    \partial_{(a}V_{b)} &= -\frac{1}{r^2}\epsilon_a\epsilon_b \equiv -\frac{1}{2}V_aV_b
\end{aligned}
\eeq
Hence after substituting $r = r_H(1 + \epsilon)$ and expanding in powers of $\epsilon$, we can can write \eqref{deltaabcd} to leading power in $\epsilon$ as
\beq\label{explicitDeltaabcd}
\Big[\frac{1}{2}\mu^2\eta_{ab}\epsilon_c\epsilon_d + \frac{1}{2}\mu^2(\lambda-1)(\eta_{ab}\eta_{cd} - \eta_{ac}\eta_{bd})\Big]\Delta^{cdef} = \delta^{ef}_{ab}\delta^{(2)}(x^a - x'^a)
\eeq
Our ansatz for $\Delta^{cdef}$ is now as follows:
\beq
\Delta^{cdef} = Q \eta^{c(e}\eta^{f)d} + T \eta^{cd}\eta^{ef} + P\eta^{cd}\epsilon^{e}\epsilon^{f}
\eeq
Substituting this ansatz in \eqref{explicitDeltaabcd}, we obtain
\beq\begin{aligned}
&\frac{1}{2}\mu^2Q(\lambda-1)(\eta_{ab}\eta^{ef}-\delta^{ef}_{ab}) + \frac{1}{2}\mu^2T(\lambda-1)\eta_{ab}\eta^{ef} + \frac{1}{2}\mu^2P(\lambda-1)\eta_{ab}\epsilon^e\epsilon^f \\&+\frac{1}{2}\mu^2Q\eta_{ab}\epsilon^e\epsilon^f + \mu^2T\eta_{ab}\eta^{ef} + \mu^2P\eta_{ab}\epsilon^e\epsilon^f = \delta^{ef}_{ab}    
\end{aligned}
\eeq
Solving this, gives
\begin{subequations}
    \begin{align}
        Q &= -\frac{2}{\mu^2(\lambda-1)} \\
        T &= \frac{2}{\mu^2(\lambda+1)} \\
        P &= \frac{2}{\mu^2(\lambda+1)(\lambda-1)}
    \end{align}
\end{subequations}
Thus the final expression for $\Delta^{cdef}$ is
\beq
\Delta^{cdef} = -\frac{2}{\mu^2(\lambda-1)}\eta^{c(e}\eta^{f)d} + \frac{2}{\mu^2(\lambda+1)}\eta^{cd}\eta^{ef} + \frac{2}{\mu^2(\lambda+1)(\lambda-1)}\eta^{cd}\epsilon^e\epsilon^f
\eeq
It can be checked that if the remaining two indices were contracted in \eqref{deltaabcd} as
\beq
\Delta^{-1}_{cdab}\Delta_0^{cdef} = \delta^{ef}_{ab}\delta^{(2)}(x-x')
\eeq
then following a similar derivation as above, the expression for $\Delta_0^{cdef}$ can be evaluated as
\beq
\Delta_0^{cdef} = -\frac{2}{\mu^2(\lambda-1)} \eta^{c(e}\eta^{f)d} + \frac{2}{\mu^2(\lambda+1)}\eta^{cd}\eta^{ef} + \frac{2}{\mu^2(\lambda-1)(\lambda+1)}\epsilon^c\epsilon^d\eta^{ef}
\eeq
Since $\eta^{c(e}\eta^{f)d}$ and $\eta^{cd}\eta^{ef}$ are symmetric under $cd \leftrightarrow ef$, we have managed to show that $\Delta_0^{cdef} = \Delta^{efcd}$. This implies that
\beq
\Delta^{-1}_{cdab}\Delta^{efcd} = \delta^{ef}_{ab}\delta^{(2)}(x-x')
\eeq
Next, we will show that we can write down an expression for $\mathcal{P}$ by solving the following propagator equations in momentum space
\begin{subequations}\label{propeqn}
    \begin{align}\label{propeqn1}
        \Delta^{-1}_{abcd}\mathcal{P}^{cd}_R + \Delta^{-1}_{L,ab}\mathcal{P} &= 0 \\
        \label{propeqn2}
    \Delta^{-1}_{R,ab}\mathcal{P}^{ab}_R + \Delta^{-1}\mathcal{P} &= 1
    \end{align}
\end{subequations}
Multiplying both the sides of \eqref{propeqn1} with $\Delta^{efab}$, we get
\begin{subequations}
    \begin{align}
      \Delta^{-1}_{abcd}\Delta^{efab}\mathcal{P}^{cd}_R &= -\Delta^{-1}_{L,ab}\Delta^{efab}\mathcal{P} \\
      \implies \delta^{ef}_{cd}\mathcal{P}^{cd}_R &= -\Delta^{-1}_{L,ab}\Delta^{efab}\mathcal{P} \\
      \implies \mathcal{P}^{ab}_R &= - \Delta^{-1}_{L,cd}\Delta^{abcd}\mathcal{P} =  - \Delta^{abcd}\Delta^{-1}_{L,cd}\mathcal{P}
    \end{align}
\end{subequations}
Substituting this back in \eqref{propeqn2}, we can write
\begin{subequations}
    \begin{align}\label{prop}
        -\Delta^{-1}_{R,ab}\Delta^{abcd}\Delta^{-1}_{L,cd}\mathcal{P} + \Delta^{-1}\mathcal{P} &= 1 \\
        \label{Kpropagator}
        \implies \mathcal{P} = \frac{1}{\Delta^{-1}-\Delta^{-1}_{R,ab}\Delta^{abcd}\Delta^{-1}_{L,cd}}
    \end{align}
\end{subequations}
In momentum space, we can evaluate using the following expressions for $\Delta^{-1}_{ab}$ and $\Delta^{abcd}$
\begin{subequations}
    \begin{align}
        \Delta^{-1} &= k^2 \\
        \Delta^{-1}_{R, ab} &= \eta_{ab}(k^2+\frac{1}{2}\mu^2(\lambda-3)) - k_ak_b \\
        \Delta^{-1}_{L ,cd} &= \eta_{cd}(k^2+\frac{1}{2}\mu^2(\lambda-1)) - k_ck_d \\
        \Delta^{abcd} &= -\frac{2}{\mu^2(\lambda-1)} \eta^{a(c}\eta^{d)b} + \frac{2}{\mu^2(\lambda+1)}\eta^{ab}\eta^{cd} + \frac{2}{\mu^2(\lambda-1)(\lambda+1)}\eta^{ab}\epsilon^c\epsilon^d
    \end{align}
\end{subequations}
Using the following identities
\begin{subequations}
    \begin{align}
        \eta^{a(c}\eta^{d)b}\eta_{cd} &= \eta^{ab} \\
        \eta^{a(c}\eta^{d)b}k_ck_d &= k^ak^b \\
        \epsilon^a\epsilon^b\eta_{ab} &= \epsilon^2 = 2 \\
        (\epsilon\cdot k)^2 = (\epsilon^ak_a)^2 &= -2\mu^2
    \end{align}
\end{subequations}
The proof for the last identity is as follows. We had shown earlier that $\partial_aV_b = -\frac{1}{r^2}\epsilon_a\epsilon_b$. In momentum space, this may be simplified by setting $r=r_H$ to obtain
\begin{subequations}
    \begin{align}
        k_a\epsilon_b &= -\frac{i}{\sqrt{2}}\mu\epsilon_a\epsilon_b \\
        \implies \epsilon^ak_a &= \eta^{ab}\epsilon_bk_a = -\frac{i}{\sqrt{2}}\mu\epsilon^2 \\
        \implies  \epsilon^ak_a &= -\sqrt{2}i\mu \\
        \implies (\epsilon\cdot k)^2 &= -2\mu^2
    \end{align}
\end{subequations}
Substituting the given expressions for $\Delta^{-1}_{ab}$ and $\Delta^{abcd}$ in \eqref{Kpropagator} and using the above identities, it can be shown, after some algebra, that the momentum space propagator $\mathcal{P}$ becomes
\beq
\mathcal{P}(k) = - \frac{\lambda+1}{\lambda-3}\frac{1}{k^2 + \mu^2\Big(\frac{\lambda^2-2\lambda+5}{\lambda-1}\Big)}
\eeq
Next, we would like to write down the form of the propagator in position space. Let us first define the propagator mass in flat Minkowski spacetime $M_{\text{Mink}}(\lambda)$ by
\beq
M_{\text{Mink}}^2(\lambda) = \mu^2\frac{\lambda^2-2\lambda+5}{\lambda-1}
\eeq
Next, we do a Fourier transform similar to the case of the scalar propagator to obtain the following expression in position space
\beq
\mathcal{P}(s) = \frac{\lambda+1}{\lambda-3}\frac{1}{2\pi i}K_0(M_{\text{Mink}}s)
\eeq

\bibliographystyle{JHEP}

\begin{thebibliography}{99}

\bibitem{Parikh:1999mf}
M.~K.~Parikh and F.~Wilczek,
\emph{{Hawking radiation as tunneling}}, \href{https://journals.aps.org/prl/abstract/10.1103/PhysRevLett.85.5042}{\emph{Phys. Rev. Lett.} {\bf 85}, 5042-5045 (2000)}, [\href{https://arxiv.org/abs/hep-th/9907001}{{\tt hep-th/9907001}}].

\bibitem{Parikh:2004ih}
M.~K.~Parikh,
\emph{{A Secret tunnel through the horizon}}, \href{https://www.worldscientific.com/doi/abs/10.1142/S0218271804006498}{\emph{Int. J. Mod. Phys. D} {\bf 13}, 2351-2354 (2004)}, [\href{https://arxiv.org/abs/hep-th/0405160}{{\tt hep-th/0405160}}].

\bibitem{Gaddam:2020mwe}
N.~Gaddam and N.~Groenenboom, \emph{{Soft graviton exchange and the information paradox}}, \href{https://journals.aps.org/prd/abstract/10.1103/PhysRevD.109.026007}{\emph{Phys. Rev. D} {\bf 109}, 026007 (2024)}, [\href{https://arxiv.org/abs/2012.02355}{{\tt
  2012.02355}}]. 

\bibitem{Gaddam:2020rxb}
N.~Gaddam, N.~Groenenboom and G.~'t Hooft, \emph{{Quantum gravity on the black hole horizon}}, \href{https://link.springer.com/article/10.1007/JHEP01(2022)023}{\emph{JHEP} {\bf 01}, 023 (2022)}, [\href{https://arxiv.org/abs/2012.02357}{{\tt
  2012.02357}}].

\bibitem{Parikh:1997ma}
M.~Parikh and F.~Wilczek,
\emph{{An Action for black hole membranes}}, \href{https://journals.aps.org/prd/abstract/10.1103/PhysRevD.58.064011}{\emph{Phys. Rev. D} {\bf 58}, 064011 (1998)}, [\href{https://arxiv.org/abs/gr-qc/9712077}{{\tt hep-th/9712077}}].

\bibitem{Kallosh:2021ors}
R.~Kallosh and A.~A.~Rahman, \emph{{Quantization of gravity in the black hole background}}, \href{https://journals.aps.org/prd/abstract/10.1103/PhysRevD.104.086008}{\emph{Phys. Rev. D} {\bf 104}, 086008 (2021)}, [\href{https://arxiv.org/abs/2106.01966}{{\tt
  2106.01966}}].

\bibitem{Kallosh:2021uxa}
R.~Kallosh, \emph{{Quantization of gravity in spherical harmonic basis}}, \href{https://journals.aps.org/prd/abstract/10.1103/PhysRevD.104.086023}{\emph{Phys. Rev. D} {\bf 104}, 086023 (2021)}, [\href{https://arxiv.org/abs/2107.02099}{{\tt
  2107.02099}}].

\bibitem{Regge:1957td}
T.~Regge and J.~A.~Wheeler, \emph{{Stability of a Schwarzschild singularity}}, \href{https://journals.aps.org/pr/abstract/10.1103/PhysRev.108.1063}{\emph{Phys. Rev. D} {\bf 108}, 1063-1069 (1957)}.

\bibitem{Zerilli:1970wzz}
F.~J.~Zerilli, \emph{{Gravitational field of a particle falling in a Schwarzschild geometry analyzed in tensor harmonics}}, \href{https://journals.aps.org/pr/abstract/10.1103/PhysRev.108.1063}{\emph{Phys. Rev. D} {\bf 2}, 2141-2160 (1970)}.

\bibitem{Candelas:1980zt}
P.~Candelas,
\emph{{Vacuum Polarization in Schwarzschild Space-Time}}, \href{https://journals.aps.org/prd/abstract/10.1103/PhysRevD.21.2185}{\emph{Phys. Rev. D} {\bf 21}, 2185-2202 (1980)}.

\bibitem{Marolf:2003bb}
D.~Marolf, \emph{{On the quantum width of a black hole horizon}},
  \href{https://link.springer.com/chapter/10.1007/3-540-26798-0_9}{\emph{Springer Proc. Phys.} {\bf 98}, 99-112 (2005)}, [\href{https://arxiv.org/abs/hep-th/0312059}{{\tt
  hep-th/0312059}}].

\bibitem{Bousso:2023kdj}
R.~Bousso and G.~Penington, \emph{{Islands far outside the horizon}}, \href{https://link.springer.com/article/10.1007/JHEP11(2024)164}{\emph{JHEP} {\bf 11}, 164 (2024)}, [\href{https://arxiv.org/abs/2312.03078}{{\tt
  2312.03078}}].

\bibitem{Banks:2024imv}
T.~Banks, P.~Draper and M.~Karydas, \emph{{Breakdown of field theory in near-horizon regions}}, \href{https://link.springer.com/article/10.1007/JHEP06(2024)153}{\emph{JHEP} {\bf 06}, 153 (2024)}, [\href{https://arxiv.org/abs/2401.03572}{{\tt
  2401.03572}}].  
  
\bibitem{Gaddam:2022pnb}
N.~Gaddam and N.~Groenenboom, \emph{{A toolbox for black hole scattering}}, [\href{https://arxiv.org/abs/2207.11277}{{\tt
  2207.11277}}].

\bibitem{Martel:2005ir}
K.~Martel and E.~Poisson, \emph{{Gravitational perturbations of the Schwarzschild spacetime: A Practical covariant and gauge-invariant formalism}}, \href{https://journals.aps.org/prd/abstract/10.1103/PhysRevD.71.104003}{\emph{Phys. Rev. D} {\bf 71}, 104003 (2005)}, [\href{https://arxiv.org/abs/gr-qc/0502028}{{\tt gr-qc/0502028}}].




\bibitem{Gukov:2022oed}
S.~Gukov, V.~S.~H.~Lee and K.~M.~Zurek, \emph{{Near-horizon quantum dynamics of 4D Einstein gravity from 2D Jackiw-Teitelboim gravity}}, \href{https://journals.aps.org/prd/abstract/10.1103/PhysRevD.107.016004}{\emph{Phys. Rev. D} {\bf 107}, 016004 (2023)}, [\href{https://arxiv.org/abs/2205.02233}{{\tt 2205.02233}}].






\bibitem{Freivogel:2024ulb}
B.~Freivogel and T.~Li, \emph{{Estimating Quantum Gravity Corrections to Correlators near Black Holes}}, [\href{https://arxiv.org/abs/2405.17570}{{\tt 2405.17570}}].







\end{thebibliography}

\end{document}